\documentclass[journal]{IEEEtran}
\ifCLASSINFOpdf
  \usepackage[pdftex]{graphicx}
\else
\fi
\usepackage{booktabs}
\usepackage{multirow}

\usepackage{amsmath}
\usepackage{amssymb}
\usepackage{bm}
\usepackage{xcolor}

\begin{document}
%
% paper title
% Titles are generally capitalized except for words such as a, an, and, as,
% at, but, by, for, in, nor, of, on, or, the, to and up, which are usually
% not capitalized unless they are the first or last word of the title.
% Linebreaks \\ can be used within to get better formatting as desired.
% Do not put math or special symbols in the title.
\title{Frequency-Sliding Generalized Cross-Correlation: A Sub-band Time Delay Estimation Approach}
%
%
% author names and IEEE memberships
% note positions of commas and nonbreaking spaces ( ~ ) LaTeX will not break
% a structure at a ~ so this keeps an author's name from being broken across
% two lines.
% use \thanks{} to gain access to the first footnote area
% a separate \thanks must be used for each paragraph as LaTeX2e's \thanks
% was not built to handle multiple paragraphs
%

\author{Maximo~Cobos,~\IEEEmembership{Senior~Member,~IEEE,}
        Fabio~Antonacci,~\IEEEmembership{Senior~Member,~IEEE,}
        Luca~Comanducci,~\IEEEmembership{Student Member,~IEEE,}
        and~Augusto~Sarti,~\IEEEmembership{Senior~Member,~IEEE}% <-this % stops a space
\thanks{M. Cobos is with the Departament d'Inform\`atica, Universitat de Val\`encia, Burjassot, 46100 Spain e-mail: (maximo.cobos@uv.es).}% <-this % stops a space
\thanks{F. Antonacci, A. Sarti and L. Comanducci are with the Dipartimento di Elettronica, Informazione e Bioingegneria
at the Politecnico di Milano, Italy. E-mail: (firstname.lastname@polimi.it)}% <-this % stops a space
\thanks{This work has been partially supported by FEDER and the Spanish Ministry of Science, Innovation and Universities under Grants RTI2018-097045-B-C21 and PRX19/00075.}}
%\thanks{Manuscript received April 19, 2005; revised August 26, 2015.}}

% note the % following the last \IEEEmembership and also \thanks - 
% these prevent an unwanted space from occurring between the last author name
% and the end of the author line. i.e., if you had this:
% 
% \author{....lastname \thanks{...} \thanks{...} }
%                     ^------------^------------^----Do not want these spaces!
%
% a space would be appended to the last name and could cause every name on that
% line to be shifted left slightly. This is one of those "LaTeX things". For
% instance, "\textbf{A} \textbf{B}" will typeset as "A B" not "AB". To get
% "AB" then you have to do: "\textbf{A}\textbf{B}"
% \thanks is no different in this regard, so shield the last } of each \thanks
% that ends a line with a % and do not let a space in before the next \thanks.
% Spaces after \IEEEmembership other than the last one are OK (and needed) as
% you are supposed to have spaces between the names. For what it is worth,
% this is a minor point as most people would not even notice if the said evil
% space somehow managed to creep in.

% The paper headers
%\markboth{Journal of \LaTeX\ Class Files,~Vol.~XX, No.~XX, September~2019}%
%{Shell \MakeLowercase{\textit{et al.}}: Bare Demo of IEEEtran.cls for IEEE Journals}

\markboth{IEEE/ACM TRANSACTIONS ON AUDIO, SPEECH, AND LANGUAGE PROCESSING}%
{Cobos \MakeLowercase{\textit{et al.}}: Bare Demo of IEEEtran.cls for IEEE Journals}

% The only time the second header will appear is for the odd numbered pages
% after the title page when using the twoside option.
% 
% *** Note that you probably will NOT want to include the author's ***
% *** name in the headers of peer review papers.                   ***
% You can use \ifCLASSOPTIONpeerreview for conditional compilation here if
% you desire.

% If you want to put a publisher's ID mark on the page you can do it like
% this:
%\IEEEpubid{0000--0000/00\$00.00~\copyright~2015 IEEE}
% Remember, if you use this you must call \IEEEpubidadjcol in the second
% column for its text to clear the IEEEpubid mark.

% use for special paper notices
%\IEEEspecialpapernotice{(Invited Paper)}

%\IEEEoverridecommandlockouts
%\IEEEpubid{\makebox[\columnwidth]{Copyright
%\copyright2019 IEEE\hfill} \hspace{\columnsep}\makebox[\columnwidth]{ }}

% COPYRIGHT NOTICE
\IEEEoverridecommandlockouts
\IEEEpubid{\begin{minipage}[t]{\textwidth}\ \\[10pt]
        \centering\footnotesize{\copyright 2019 IEEE.  Personal use of this material is permitted.  Permission from IEEE must be obtained for all other uses, in any current or future media, including reprinting/republishing this material for advertising or promotional purposes, creating new collective works, for resale or redistribution to servers or lists, or reuse of any copyrighted component of this work in other works.}
\end{minipage}} 

% make the title area
\maketitle

% As a general rule, do not put math, special symbols or citations
% in the abstract or keywords.
\begin{abstract}
The generalized cross-correlation (GCC) is regarded as the most popular approach for estimating the time difference of arrival (TDOA) between the signals received at two sensors. Time delay estimates are obtained by maximizing the GCC output, where the direct-path delay is usually observed as a prominent peak. Moreover, GCCs play also an important role in steered response power (SRP) localization algorithms, where the SRP functional can be written as an accumulation of the GCCs computed from multiple sensor pairs. Unfortunately, the accuracy of TDOA estimates is affected by multiple factors, including noise, reverberation and signal bandwidth. In this paper, a sub-band approach for time delay estimation aimed at improving the performance of the conventional GCC is presented. The proposed method is based on the extraction of multiple GCCs corresponding to different frequency bands of the cross-power spectrum phase in a sliding-window fashion. The major contributions of this paper include: 1) a sub-band GCC representation of the cross-power spectrum phase that, despite having a reduced temporal resolution, provides a more suitable representation for estimating the true TDOA; 2) such matrix representation is shown to be rank one in the ideal noiseless case, a property that is exploited in more adverse scenarios to obtain a more robust and accurate GCC; 3) we propose a set of low-rank approximation alternatives for processing the sub-band GCC matrix, leading to better TDOA estimates and source localization performance. An extensive set of experiments is presented to demonstrate the validity of the proposed approach.
\end{abstract}

% Note that keywords are not normally used for peerreview papers.

% For peer review papers, you can put extra information on the cover
% page as needed:
% \ifCLASSOPTIONpeerreview
% \begin{center} \bfseries EDICS Category: 3-BBND \end{center}
% \fi
%
% For peerreview papers, this IEEEtran command inserts a page break and
% creates the second title. It will be ignored for other modes.
\IEEEpeerreviewmaketitle

\section{Introduction}
\IEEEPARstart{T}{ime} delay estimation (TDE) refers to finding the time differences-of-arrival (TDOAs) between signals received at an array of sensors. Traditionally, TDE has played an important role in many location-aware systems, including radar, sonar, wireless systems, sensor calibration or seismology. In acoustic signal processing, TDE is essential for localizing and tracking acoustic sources \cite{CobosWCMC2017, Evers2019}. Automatic camera steering \cite{Marti2011}, speaker localization \cite{Belloch2015} or speech enhancement systems are application examples strongly relying on TDE \cite{Chen2006}.

The generalized cross-correlation (GCC), originally proposed by Knapp and Carter in their 1976 seminal paper \cite{KnappCarter1976}, is still today the most popular technique for TDE. By using GCCs, the TDOA between two signals is estimated as the time lag that maximizes the cross-correlation between filtered versions of such signals. In this context, the GCC may consider different weighting functions (filters) characterized by a particular behavior \cite{Ianniello82, WeinsteinI, WeinsteinII}. Roth, smoothed coherence transform, Eckart or Hannan-Thomson (maximum likelihood) are examples of such weighting schemes \cite{KnappCarter1976, Carter1987}. 
\textcolor{black}{The GCC with phase transform (GCC-PHAT) has been repeatedly shown to be a suitable alternative for TDE in real reverberant scenarios \cite{Gustafsson2003, Perez2012}. However,  it has also been demonstrated that its performance can only be considered optimal under uncorrelated noise conditions and a high signal-to-noise ratio (SNR) \cite{Zhang2008}}. The present work is particularly focused on PHAT weighting, thus, the terms GCC and GCC-PHAT will be used indistinctly throughout this paper. Since the GCC was first proposed, many approaches have appeared to ameliorate the robustness of TDE techniques, with improvements that are mostly achieved by exploiting the spatial diversity provided by more than two microphones \cite{Chen2003, Alameda2012,BenestyChen2004}, blind channel estimation \cite{BenestyAED, Doclo2003}, modified GCC weightings \cite{Padois2019}, or the incorporation of some a priori information \cite{Brandstein1997, Yegnanarayana2005}. \textcolor{black}{Further improvements to deal with reverberation and correlated noise fields have also been proposed \cite{dvorkind2005time}, addressing some of the limitations of the GCC. In addition, other  methods  exploiting temporal diversity by means of Bayesian filtering have shown great potential \cite{Vermaak2001, Wu2017}.} 

In contrast to the above methods, this paper proposes a novel sub-band approach to TDE with two sensors which is not directly categorized within the above processing improvements. Concretely, the method is based on the exploration of the cross-power spectrum phase by following a sliding window approach, obtaining a set of sub-band GCCs that encode the contribution of different frequency bands to the estimated TDOA. The resulting sub-band GCC matrix is shown to be rank-one for a full-band signal in the noiseless single-path case. This fact is exploited to obtain a robust GCC in adverse scenarios, proposing low-rank approximations of the sub-band GCC matrix that ultimately lead to better estimation accuracy, reduced level of spurious peaks and lower probability of anomalous estimates.

The rest of the paper is structured as follows. Section~\ref{sec:tde} summarizes the background concerning GCC-based TDOA estimation. Section \ref{sec:FS_GCC} presents the frequency-sliding GCC (FS-GCC) representation. Section~\ref{sec:low_rank} discusses our proposed methods for TDE based on the FS-GCC. The experimental evaluation is in Section~\ref{sec:experiments}. Finally, the conclusions of this work are summarized in Section~\ref{sec:conclusion}.

\section{Time Delay Estimation}
\label{sec:tde}

This section summarizes the conventional GCC approach for TDE. To this end, the ideal anechoic model is first presented, discussing the main problems arising in realistic acoustic conditions. 

\subsection{Anechoic Signal Model}

Let us consider a pair of sensors with spatial coordinates given by column vectors $\mathbf{m}_1$,  $\mathbf{m}_2$ $\in \mathbb{R}^{3}$ and an emitting acoustic source located at $\mathbf{s} \in \mathbb{R}^{3}$. The time difference-of-arrival (TDOA) measured in samples is defined as
\begin{equation}
    \tau_0 \triangleq \left\lfloor  \frac{\|\mathbf{s} -\mathbf{m}_1\| -  \|\mathbf{s} -\mathbf{m}_2\|}{c} f_s\right\rceil = \eta_1 - \eta_2,
    \label{eq:tdoa}
\end{equation}
where $c$ is the wave propagation speed, $\left\lfloor \cdot \right \rceil$ denotes the rounding operator and $f_s$ is the sampling frequency. The terms $\eta_1$ and $\eta_2$ represent the time of flight (TOF) of the sound to the sensors in samples. 

Assuming an anechoic scenario, the signals received by the two sensors can be modeled as
\begin{equation}
    x_m[n] = \beta_m s[n-\eta_m] + w_m[n], \quad m=1,2,
\end{equation}
where $\beta_m \in  \mathbb{R}_{+}$ is a positive amplitude decay factor, $s[n]$ is the source signal and $w_m[n]$ is an additive noise term. In the discrete-time Fourier transform (DTFT) domain, the sensor signals can be written as
\begin{equation}
    X_m(\omega) = \beta_m S(\omega)e^{-j\omega \eta_m} + W_m(\omega),  \quad m=1,2,
\end{equation}
where $S(\omega)$, $W_m(\omega) \in \mathbb{C}$, are the DTFTs of the source signal and the noise signal, respectively, and $j = \sqrt{-1}$.

%The generalized cross-correlation with phase transform (GCC-PHAT) is a widely used method in acoustic signal processing for estimating the time-difference of arrival (TDOA) corresponding to a pair of sensors. The TDOA is given by
%\begin{equation}
    %\tau_{12} = \eta_1 - \eta_2 =  \frac{\|\mathbf{x}_s -\mathbf{m}_1\|}{c} -  \frac{\|\mathbf{x}_s -\mathbf{m}_2\|}{c},
%\end{equation}
%where the terms $\eta_1$ and $\eta_2$ represent the time of flight (TOF) of the sound to the first and second sensors with sound propagation speed $c$. The vector  $\mathbf{x}_s$ contains the coordinates of the source, while the vectors $\mathbf{m}_1$ and $\mathbf{m}_2$ contain the coordinates of the first and second sensors, respectively.

%Assuming an anechoic scenario, the signals of the two sensors can be modeled as
%\begin{equation}
%    x_m(t) = \alpha_m s(t-\eta_m) + n_m(t), \quad m=1,2,
%\end{equation}
%where $\alpha_m$ is a positive amplitude decay factor, $s(t)$ is the source signal and $n_m(t)$ is an additive noise term. In the discrete-time Fourier transform (DTFT) domain, the sensor signals can be written as
%\begin{equation}
%    X_m(\omega) = \alpha_m S(\omega)e^{-j\omega \eta_m} + N_m(\omega),  \quad m=1,2,
%\end{equation}
%where $S(\omega)$ and $N_m(\omega)$ are the DTFTs of the source signal and the noise signal, respectively.

\subsection{Generalized Cross-Correlation}

The GCC of a pair of sensor signals is defined as the inverse Fourier transform of the weighted cross-power spectrum, i.e.
\begin{equation}
    R[\tau] \triangleq \frac{1}{2\pi}\int_{-\pi}^{\pi}\Psi(\omega)e^{j\omega \tau} d\omega =\mathcal{F}^{-1}\left\{ \Psi(\omega) \right\},
\end{equation}
where $\tau \in \mathbb{Z}$ represents time delay and $\Psi(\omega) \in \mathbb{C}$ represents the phase transform (PHAT) cross-power spectrum:
\begin{equation}
    \Psi(\omega) \triangleq \frac{X_1(\omega)X^*_2(\omega)}{\left|X_1(\omega)X^*_2(\omega)\right|},
\end{equation}
where $(\cdot)^{*}$ denotes complex conjugation. Note that, in the ideal anechoic and noiseless case, 
\begin{equation}
    \Psi(\omega) = \frac{\beta_1 \beta_2 |S(\omega)|^2}{\left|\beta_1 \beta_2 |S(\omega)|^2\right|}e^{-j\omega\tau_0} = e^{-j\omega\tau_0},
    \label{eq:full_band}
\end{equation}
leading to
\begin{equation}
    R[\tau] = \mathcal{F}^{-1}\left\{e^{-j\omega\tau_0} \right\} = \delta[\tau-\tau_0],
    \label{eq:anegcc}
\end{equation}
where $\delta[\tau]$ is the Kronecker delta function. Therefore, the ideal GCC-PHAT for a full-band source signal will show a unit impulse located at the true TDOA. This motivates the use of the following estimator for time-delay estimation over signals acquired by a pair of sensors:
\begin{equation}
    \hat{\tau}_0 = \arg\max_\tau R[\tau].
\end{equation}

The GCC-PHAT inherently discards the magnitude information of the signals and provides a time-delay estimate based only on the phase of the cross-power spectrum, i.e. the phase of the frequency domain analysis of the cross-correlation between the two signals. It is important to remark that the measured time delay is an integer multiple of the sampling period. Note, however, that a finer resolution can be achieved by interpolating between consecutive samples of the GCC function if necessary. 

\subsection{Problems in Realistic Scenarios}

It is well-known that several problems arise when using GCC-PHAT for estimating the TDOA in realistic scenarios. Indeed, TDOA measurements are very sensitive to reverberation, noise, and the presence of potential interferers:
\begin{itemize}
    \item In reverberant environments, for certain locations and orientations of the source signal, the peak of the GCC related to a reflective path could overcome that of the direct path.
    \item In noisy scenarios, for some time instants, the noise level could exceed that of the signal, making the estimated TDOA unreliable.
    \item Peaks corresponding to the direct path, reflections or combinations of interfering signals may also lead to errors when estimating the TDOA of a target source.
\end{itemize}

\begin{figure*}[!t]
\includegraphics[width=\textwidth]{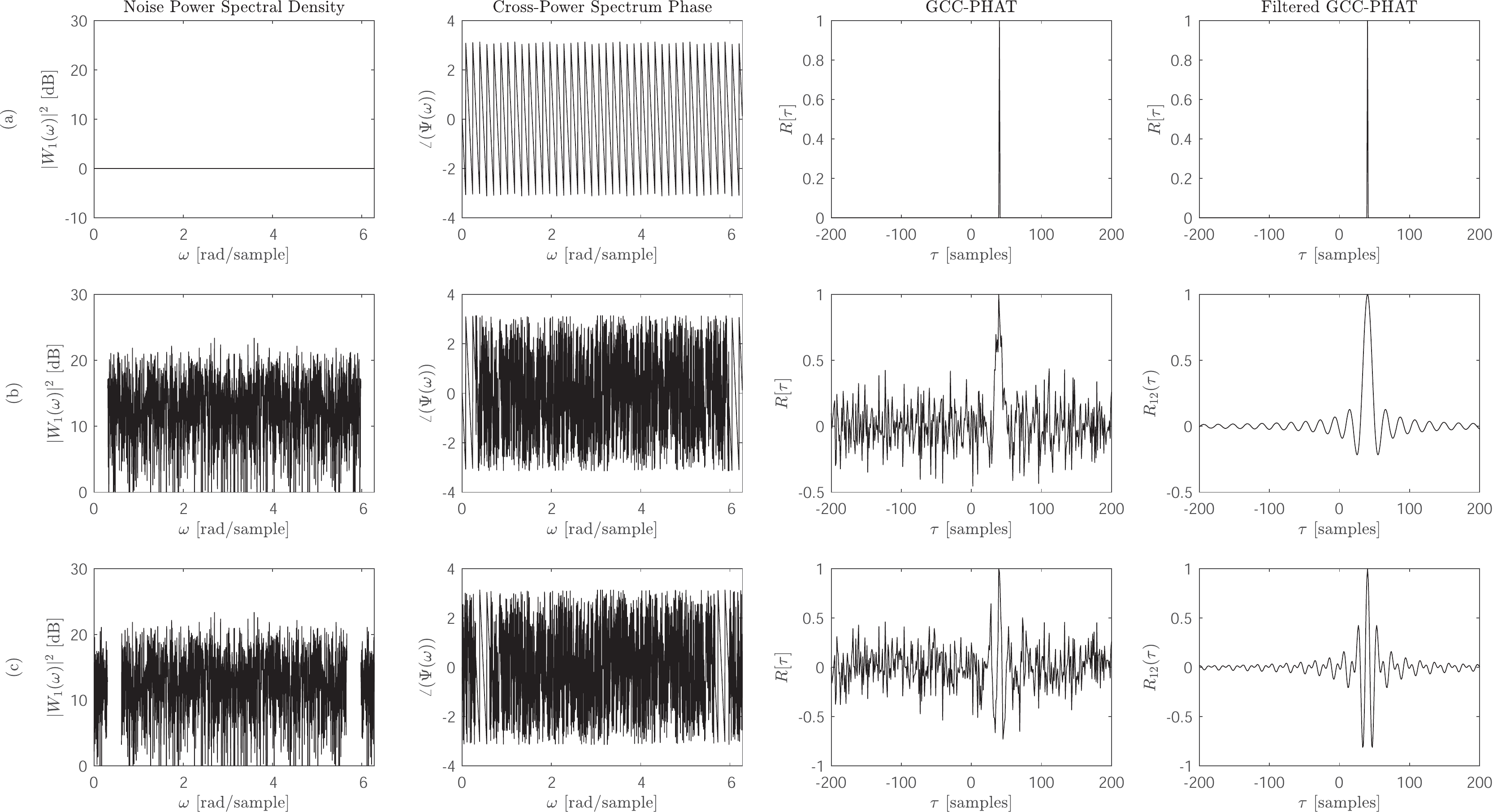}
\caption{\textcolor{black}{GCCs obtained from two signals corrupted by additive white noise ($\textrm{SNR}=-15$ dB) affecting different frequency bands: (a) No noise. (b) Noise affecting high frequencies. (c) Noise affecting high and low frequencies. The true TDOA is $\tau_{0}=40$ samples.}}
\label{fig:theo_GCC}
\end{figure*}

\textcolor{black}{The above issues can be even more problematic when the spectral characteristics of the target source or the additive noise lead to a reduced signal-to-noise ratio (SNR) at some frequency bands \cite{CobosSPL2017}. The phase information can be completely lost at those frequency bands where either there is no signal information or where its content is especially affected by the noise. Consider, for example, the (normalized) GCCs shown in Figure \ref{fig:theo_GCC}. The top row corresponds to the ideal case of a noiseless full-band signal. The other two rows correspond to the same signal but affected by different spectral noise profiles. As it can be observed, the noisy phase introduces many spurious peaks that can cause anomalous TDOA estimates. Although noisy frequency bands could be filtered out to obtain a cleaner GCC (last column of Fig. \ref{fig:theo_GCC}), knowing beforehand the best frequency range accommodating the time delay information is not always straightforward. Moreover, the rippling GCCs resulting from pass-band-like signals (third row) complicate substantially source localization tasks in adverse conditions \cite{CobosSPL2017}.}

%\begin{figure*}[!t]
%\includegraphics[width=\textwidth]{GCCsf.pdf}
%\caption{GCCs obtained by filtering the normalized cross-power spectrum for the low-pass (first row) and pass-band (last row) signals. The true TDOA is $\tau_{12}=40$.}
%\label{fig:GCCsf}
%\end{figure*}

\section{Frequency-Sliding Generalized Cross-Correlation}
\label{sec:FS_GCC}

This section proposes a frequency-sliding GCC-PHAT method that generalizes the filtering approach discussed in the last section with the aim of obtaining a robust GCC representation. This representation will be useful to explore the likelihood of different frequency bands contributing to a direct-path delay estimate.

\subsection{Sub-band GCCs}

Let us define the sub-band GCC for an arbitrary frequency band $l$ as 
\begin{eqnarray}
    R[\tau, l] &\triangleq& \frac{1}{2\pi}\int_{-\pi}^{\pi}\Psi(\omega+\omega_l)\Phi(\omega)e^{j\omega \tau} d\omega \\ \nonumber
    &=& \mathcal{F}^{-1}\left\{ \Psi(\omega+\omega_l)\Phi(\omega) \right\},
\end{eqnarray}
where $\omega_l$ is the frequency offset corresponding to band $l$. The function $\Phi(\omega) \in \mathbb{R}$ is a symmetric frequency-domain window, centered at $\omega = 0$ with frequency support $B_\Phi~\in~[0,\pi]$, i.e. $\Phi(\omega) = 0$ for $|\omega|\geq B_\Phi$.

The effect of the window and the cross-power spectrum can be separated as follows:
\begin{eqnarray}
    R[\tau,l] &=& \mathcal{F}^{-1}\left\{ \Psi(\omega+\omega_l) \right\}\ast \mathcal{F}^{-1}\left\{\Phi(\omega) \right\} \nonumber \\ 
    &=& \left(e^{-j\omega_l\tau} \mathcal{F}^{-1}\left\{ \Psi(\omega) \right\}\right)\ast \phi[\tau] \nonumber\\ 
    &=& \left(e^{-j\omega_l\tau}R[\tau] \right)\ast \phi[\tau],
    \label{eq:sub_band_gcc}
\end{eqnarray}
or equivalently, 

\begin{eqnarray}
    R[\tau,l] &=& \mathcal{F}^{-1}\left\{ \Psi(\omega)\Phi(\omega-\omega_l) \right\}e^{-j\omega_l \tau,} \nonumber \nonumber \\ 
    &=& \left( R[\tau]\ast \left( \phi[\tau] e^{j\omega_l\tau} \right)\right) e^{-j\omega_l \tau},
    \label{eq:R12_eq}
\end{eqnarray}
where $\phi[\tau]\in \mathbb{R}$ represents the inverse Fourier transform of the spectral window $\Phi(\omega)$ and $\ast$ denotes convolution. 
%For example, if $\Phi(\omega)$ is selected to be a rectangular window, i.e. $\Phi(\omega) = \mathrm{rect}\left(\frac{\omega}{2 B_\Phi} \right)$, the sub-band GCC would be given by
%\begin{equation}
%     R[\tau,l] = \left(e^{-j\omega_l\tau}R[\tau]\right) \ast \frac{\sin \left(B_\Phi \tau\right) }{\pi \tau}.
%\end{equation}

A frequency-sliding sub-band GCC can be obtained by sweeping the cross-power spectrum phase over (possibly) overlapping frequency bands:
\begin{equation}
    \omega_l = lM_\Phi, \quad l=0,\dots,L-1,
\end{equation}
where $M_\Phi$ is the frequency hop. The number of bands $L$ can be conveniently chosen to cover those frequencies up to the Nyquist limit:
\begin{equation}
    L = \left\lfloor \frac{\pi-B_\Phi+ M_\Phi}{M_\Phi} \right\rfloor.
\end{equation}

\begin{figure}[!b]
\includegraphics[width=\columnwidth]{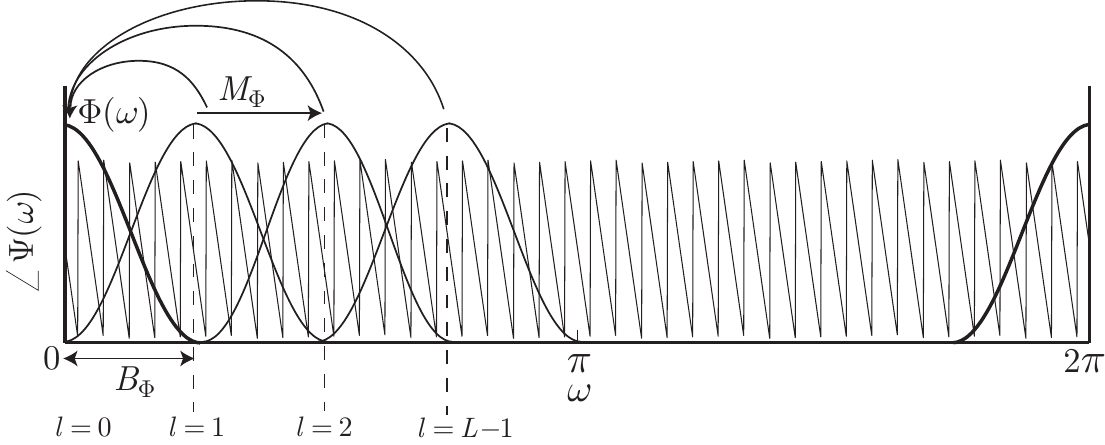}
\caption{Interpretation of the frequency-sliding GCC.}
\label{fig:sliding}
\end{figure}

An interpretation of such sliding operation is shown in Fig.~ \ref{fig:sliding}. As given by Eq.~(\ref{eq:R12_eq}), the sub-band GCCs can be interpreted as the product of $\Psi(\omega)$ with a shifted version of $\Phi(\omega)$ to $\omega_l$, shifted back in frequency to zero before taking the inverse Fourier transform. 

\subsection{Frequency-Sliding GCC Matrix}
In practice, sub-band GCCs are extracted considering the discrete Fourier transform (DFT) of the microphone signals $x_m[n]$:
\begin{equation}
    \mathbf{X}_m = \bigl[X_m[0],\, X_m[1],\dots,X_m[N-1]\bigr]^{T},\quad m=1,2,
\end{equation}
where the elements of $\mathbf{X}_{m}\in \mathbb{C}^{N}$ are the coefficients $X_m[k]$ corresponding to the discrete frequencies $\omega_k=k\frac{2\pi}{N}$. Similarly, consider the vector $\bm{\Phi}$ containing the discrete-frequency samples $\Phi[k]~=~ \Phi(\omega_k)$ of the selected spectral window \begin{equation}
    \bm{\Phi}= \bigl[\Phi[0],\, \Phi[1],\dots,0,\,0,\dots,\Phi[N-1]\bigr]^{T}\in \mathbb{R}^{N},
\end{equation}
symmetrically padded with zeros to contain only $B~=~\left\lfloor 2 B_\Phi \frac{N}{2\pi}\right\rceil$ non-zero elements.

The sub-band GCC vectors $\mathbf{r}_{l}\in \mathbb{C}^{N}$ are obtained by taking the inverse DFT of the windowed PHAT spectrum, i.e.
\begin{equation}
    \mathbf{r}_l = \bigl[r_l[0],\, r_l[1],\dots, r_l[N-1]\bigr]^{T}, \quad l=0,\dots,L-1,
\end{equation}
with
\begin{equation}
    r_l[n] = \frac{1}{N}\sum_{k=0}^{N-1}\frac{X_1[k+lM]X^*_2[k+lM]}{\bigl\lvert X_1[k+lM]X^*_2[k+lM]\bigr\rvert}\Phi[k]e^{j\frac{2\pi}{N}kn},
\end{equation}
where $M = \left\lfloor M_\Phi \frac{N}{2\pi} \right\rceil$ is the discrete frequency hop.

The frequency-sliding GCC (FS-GCC) matrix is constructed by stacking all the sub-band GCC vectors together, i.e. 
\begin{equation}
    \mathbf{R}= \left[\mathbf{r}_{0},\, \mathbf{r}_{1},\dots,\mathbf{r}_{L-1}  \right] \in \mathbb{C}^{N\times L},
\end{equation}
noting that
\begin{equation}
    \centering
\mathbf{R}[n,l] = \begin{cases}
    R[n,l], & n = 0,\dots,N/2 - 1.\\
    R[n-N,l], & n = N/2,\dots,N-1.
  \end{cases}
  \end{equation}

\subsection{Ideal Sub-band GCC Model}

To get an insight into the properties underlying the presented FS-GCC representation, let us start by considering the ideal full-band cross-power spectrum of Eq.~(\ref{eq:full_band}), $\Psi(\omega)~=~ e^{-j\omega\tau_0}$. By inserting Eq.~(\ref{eq:anegcc}) into Eq.~(\ref{eq:sub_band_gcc}), the ideal sub-band GCCs (denoted with a tilde) are given by
\begin{eqnarray}
    \tilde{R}[\tau,l] &=& e^{-j\omega_l\tau} \delta[\tau-\tau_0]\ast \phi[\tau] \\ \nonumber 
    &=& e^{-j\omega_l\tau_0}\phi[\tau-\tau_0], \quad l=1,\dots,L.
\end{eqnarray}

By analyzing the above result, some interesting observations arise:
\begin{itemize}
    \item The magnitudes of the sub-band GCCs are independent of the selected frequency offset $\omega_l$, and correspond to the inverse Fourier transform of the sub-band analysis window centered at the true TDOA:
    \begin{equation}
        \bigl\lvert \tilde{R}[\tau,l] \bigr\rvert = \bigl\lvert \phi[\tau-\tau_0]\bigr\rvert \quad \forall l.
    \end{equation}
    
    \item The maximum value of $|\tilde{R}[\tau,l]|$ is located at $\tau~=~ \tau_0$ for every sub-band $l$, $l=0,\dots,L-1$.

    \item The peak width at the true TDOA \textcolor{black}{($\mathcal{W}$)} as well as the level of its side lobes, depends on the selected type of window $\Phi$ and its frequency support $B$ \textcolor{black}{(e.g. $\mathcal{W}=2N/B$ for a rectangular window, while $\mathcal{W}=4N/B$ for a Hann window).}
    \item Since the spectral window is real and even, $\phi[\tau]$ is also real and even. Then, the phase pattern of the set of sub-band GCCs corresponds to a downsampled version of $\angle \left(\Psi(\omega)\right)$, modulated by the sign of $\phi[\tau]$.
    \item The full-band conventional GCC can be recovered from the sub-band GCCs if the constant overlap-add (COLA) property is fulfilled.
\end{itemize}
    
Consequently, the FS-GCC matrix will present the following features:

\begin{itemize}
    \item The columns of $\mathbf{R}$ correspond to the window response shifted to the true TDOA and multiplied by a different complex number, i.e.
    \begin{equation}
        \tilde{\mathbf{r}}_l = e^{-j\omega_l\tau_0}\bm{\phi}_0,
        \label{eq:perfect_column}
    \end{equation}
    where $\bm{\phi}_0 \in \mathbb{R}^{N}$ is a vector containing $N$ samples of the shifted window response:
    \begin{equation}
    \footnotesize
        \begin{split}
        \bm{\phi}_0 &\triangleq \bm{\phi}[n-\tau_0] =  \bigl[
        \begin{matrix} \phi[0-\tau_0] & \dots & \phi[N/2-1-\tau_0]  \end{matrix} \\
        &\qquad\qquad \begin{matrix} \phi[-N/2-\tau_0] & \dots & \phi[-1-\tau_0]\bigr]^{T}  \end{matrix}.
        \end{split}
        \label{eq:Phi_vector}
    \end{equation}

    \item The ideal FS-GCC matrix, $\tilde{\mathbf{R}}$, can be expressed as an outer product, resulting in a rank-one matrix:
     \begin{equation}
        \tilde{\mathbf{R}} = \bm{\phi}_0\mathbf{e}^{H},
       \label{eq:mod_ideal}
    \end{equation}
    where $\mathbf{e} \triangleq [e^{j\omega_0 \tau_0},\dots,e^{j\omega_{L-1}\tau_0}]^{T}\in \mathbb{C}^{L}$ and $(\cdot)^{H}$ denotes the conjugate transpose operator. 
    
    %The sub-band GCC matrix $\mathbf{R}_{12}$ has rank one, since all its columns are equal but multiplied by a different complex number:
    %\begin{equation}
    %    \mathbf{r}_l = e^{-j\omega_l\tau_{12}}\bm{\phi}^{(\tau_{12})},
    %\end{equation}
    %where $\bm{\phi}^{(\tau_{12})} \in \mathbb{R}^{N \times 1}$ is a vector containing $N$ samples of the shifted window response:
    %\begin{equation}
    %    \begin{split}
    %    \bm{\phi}^{(\tau_{12})} &= [
    %    \begin{matrix} \phi[0-\tau_{12}] & \dots & \phi[N/2-1-\tau_{12}]  \end{matrix} \\
    %    &\qquad\qquad \begin{matrix} \phi[-N/2-\tau_{12}] & \dots & \phi[-1-\tau_{12}]  \end{matrix} ]^{T}.
    %    \end{split}
    %\end{equation}
    %Equivalently, the GCC matrix can be expressed as
    % \begin{equation}
    %    \mathbf{R}_{12} = \bm{\phi}\mathbf{e}^{H},
    %\end{equation}
    %where $\mathbf{e} = [e^{j\omega_0 \tau_{12}},\dots,e^{j\omega_{L-1}\tau_{12}}]^{T}$ and $(\cdot)^{H}$ denotes the conjugate transpose operator.
\end{itemize}

This last observation will be used to estimate robustly the time-delay in realistic cases (noisy and/or reverberant scenarios with reduced/varying bandwidth signals).

\subsection{Noisy Sub-band GCC Model}

In a more general case, the columns of the FS-GCC matrix will contain different amounts of noise depending on the SNR characterizing the different sub-bands, i.e.
\begin{equation}
    \mathbf{r}_l = \alpha_l \tilde{\mathbf{r}}_l+(1-\alpha_l)\mathbf{n}_l, \quad l = 0,\dots,L-1,
    \label{eq:model_noisy}
\end{equation}
where $\mathbf{n}_l \in \mathbb{C}^{N}$ models the GCC noise component of the $l$-th sub-band, and $\alpha_l \in[0,1]$ is a scalar that balances the contribution of the noise in the $l$-th sub-band. Since the noise vectors in the frequency domain have a normalized magnitude spectrum given by the spectral window $\Phi(\omega)$, it follows from Parseval's theorem that
\begin{equation}
\mathbf{n}_l^{H}\mathbf{n}_l = \|\mathbf{n}_l\|^2 = \|\bm{\phi}\|^2 =  \int_{-\pi}^{\pi}\Phi^2(\omega) d\omega.
\end{equation}
Similarly, the $\mathbf{R}$ matrix can be expressed as:
\begin{equation}
    \mathbf{R} =  \tilde{\mathbf{R}}\mathbf{G} + \mathbf{N}(\mathbf{I}-\mathbf{G}),
\end{equation}
where $\mathbf{I} \in \mathbb{R}^{L \times L}$ is the identity matrix, $\mathbf{N} \in \mathbb{C}^{N\times L}$ is the matrix containing all the noise vectors, i.e.
\begin{equation}
    \mathbf{N} = \left[\mathbf{n}_{0},\, \mathbf{n}_{1},\dots,\mathbf{n}_{L-1}  \right],
\end{equation}
 and $\mathbf{G} \in \mathbb{R}^{L \times L}$ is the diagonal matrix
\begin{equation}
    \mathbf{G} =  \begin{bmatrix}
    \alpha_0 & 0 & \dots & 0 \\
    0 & \alpha_1 & \dots & 0 \\
    \vdots & \vdots & \ddots & \vdots \\
    0 & 0 & \dots & \alpha_{L-1}
  \end{bmatrix}.
\end{equation}

In contrast to the ideal case, the non-ideal FS-GCC matrix $\mathbf{R}$ will have rank $L$ unless more than one of the diagonal entries $\alpha_l$ are equal to 1.

\begin{figure}[!t]
\includegraphics[width=\columnwidth]{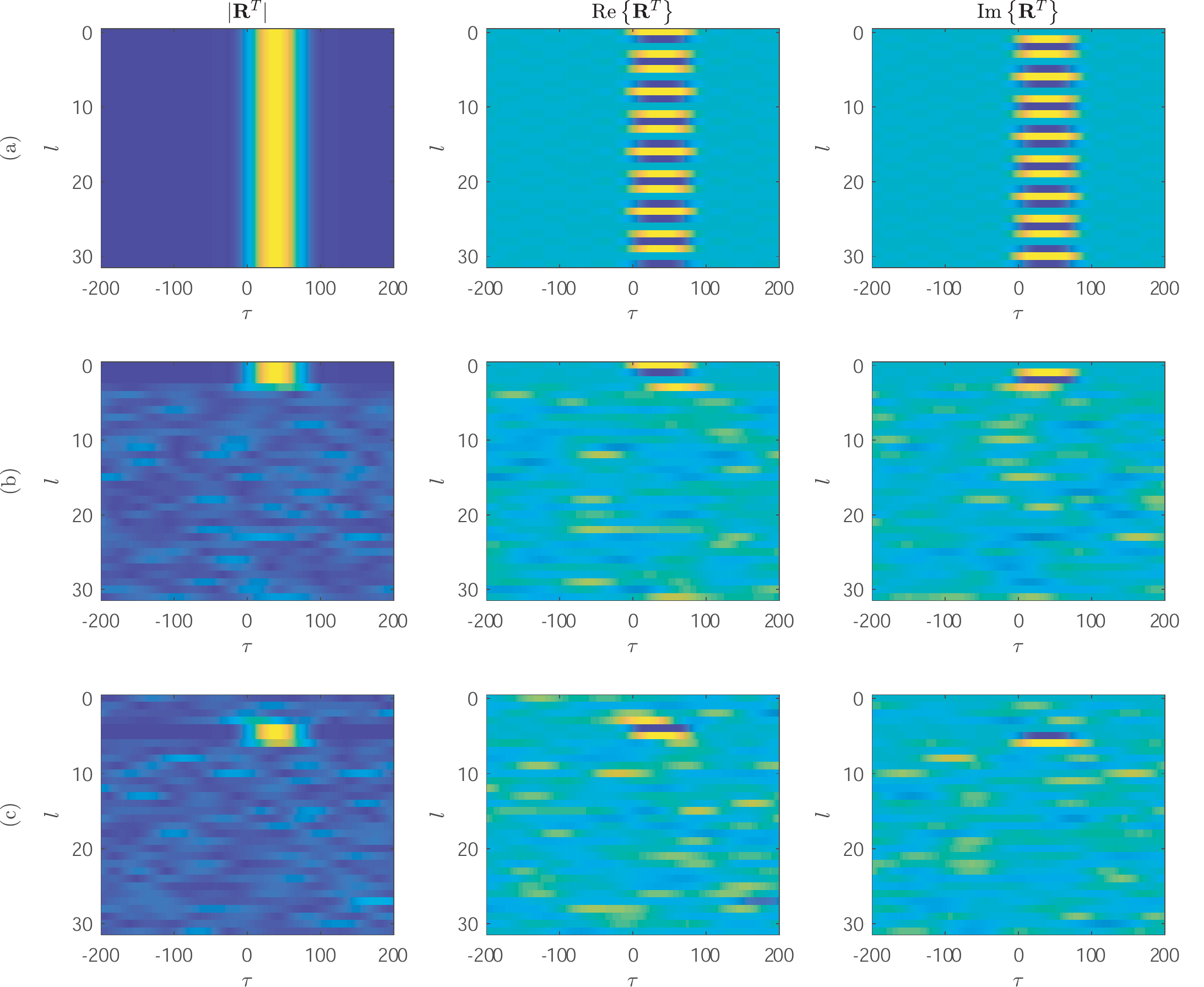}
\caption{\textcolor{black}{Sub-band GCCs obatined for the synthetic examples shown in Fig.\ref{fig:theo_GCC} ($\textrm{SNR} = -15$ dB, $\tau_0 = 40$): magnitude (left), real part (center) and imaginary part (right).}}
% \caption{Sub-band GCCs obtained for the synthetic examples shown in Fig.\ref{fig:theo_GCC} and for a real speech signal \textcolor{black}{corrupted by white noise ($\textrm{SNR} = 25$ dB)}: (a) full-band; (b) low-pass; (c) pass-band. (d) Examples for three different frames of a speech signal.}
\label{fig:GCCsb}
\end{figure}

\subsection{Examples}

Fig.~\ref{fig:GCCsb}(a-c) show the resulting FS-GCCs for the three signals analyzed in Fig.~~\ref{fig:theo_GCC}. A Hann spectral window with a frequency support of $B = 64$ bins ($B_\Phi = 0.1$) and 50\% overlap was used, resulting in $L=32$ sub-bands. As it can be observed, the first row corresponding to the \textcolor{black}{noiseless} full-band signal shows a peak perfectly aligned across all the analyzed bands and centered at the true TDOA \textcolor{black}{($\tau_0 = 40$ samples)}. Due to the use of the Hann window, the side-lobes are considerably small. The other rows corresponding to the \textcolor{black}{signal affected by noise} show also a similar pattern, but only at those sub-bands where the SNR is sufficient to preserve the phase information of the cross-power spectrum. The noise over the rest of sub-bands can be also identified, showing that no useful delay information can be extracted from other frequencies. 

\textcolor{black}{An example considering a more realistic signal can be observed in the first column of Fig. \ref{fig:GCCspeech_svd}, corresponding to three different time frames of a male speech signal ($N=2048$ samples, $f_s = 44.1$ kHz)}. In this case, $B$ has been set to 128 bins, resulting in a narrower peak. As observed, the useful bandwidth of common audio signals as speech may change considerably with time and the FS-GCC representation shows clearly which frequency bands are actively contributing to a reliable time delay estimate. Note also that the last speech frame (c) shows a bandwidth that is considerably narrower than the other two, which makes TDE more difficult in this case.

While the full-band ideal case results in a high peak continuously present over all the frequency-bands leading to a rank one matrix, in a general case, noise and reflections will interact with the linear phase component corresponding to the direct-path delay, producing a matrix with full column rank. TDE from the FS-GCC matrix should be therefore oriented towards the extraction of the reliable components of $\mathbf{R}$, discarding properly the information from noisy bands. 

\textcolor{black}{Finally, a rule of thumb to select an appropriate value for the involved parameters can be established by allowing the main lobe width ($\mathcal{W}$) of the GCC to fit within the expected range of possible time delays, i.e. $ \mathcal{W} \leq f_s \frac{d}{c}$, where $d$ is the inter-sensor distance. For example, for a Hann window ($\mathcal{W} = 4N/B$), $d=0.5$ m and $f_s = 44100$ Hz, it follows that $B \approx 128$ and $M=32$ (75\% overlap).}

\section{Low-Rank Approximations of the FS-GCC Matrix}
\label{sec:low_rank}

This section exploits the properties of the presented sub-band FS-GCC representation by proposing a low-rank approximation framework based on singular value decompositions. Such low-rank representations will be shown to provide a robust GCC for TDE in adverse conditions.

\subsection{Singular Value Decomposition}

A low-rank approximation of $\mathbf{R}$ can be obtained by solving
\begin{equation}
    \min_{\hat{\mathbf{R}}} \quad \| \mathbf{R}-\hat{\mathbf{R}}\|_F, \quad \textrm{subject to} \quad \textrm{rank}(\hat{\mathbf{R}})\leq r
\end{equation}
where $r$ is the rank of the approximating matrix $\hat{\mathbf{R}}$, and $\| \cdot \|_{F}$ denotes the Frobenius norm. The problem has analytic solution  in terms of the singular value decomposition (SVD) of $\mathbf{R}$, as given by the Eckart-Young-Mirsky theorem. Let us factorize $\mathbf{R}$ as
\begin{equation}
    \mathbf{R} = \mathbf{U}\mathbf{\Sigma}\mathbf{V}^H,
\end{equation}
where $\mathbf{U}~\in~\mathbb{C}^{N \times L}$ is the matrix containing the left singular vectors, $\mathbf{\Sigma}~\in ~\mathbb{R}^{L \times L}$ is the diagonal matrix containing the ordered singular values and $\mathbf{V}~\in~ \mathbb{C}^{L \times L}$ is the matrix containing the right singular vectors. The particular rank-$r$ matrix that best approximates $\mathbf{R}$ is given by
\begin{equation}
    \mathbf{R^{r}}=\sum_{i=1}^{r}\sigma_i\mathbf{u}_i \mathbf{v}_i^{T},
\end{equation}
where $\mathbf{u}_i$ and $\mathbf{v}_i$ are the $i$-th columns of the corresponding SVD matrices and $\sigma_i$ are the ordered singular values \cite{StrangBook2019}.

% The SVD can be interpreted as a decomposition of a matrix into a weighted, ordered sum of separable matrices as follows:
% \begin{equation}
%     \mathbf{R} = \sum_{i}\mathbf{R}^{(i)} = \sum_{i}\sigma_i \mathbf{u}_i \mathbf{v}_i^{T},
% \end{equation}
% where $\mathbf{u}_i$ and $\mathbf{v}_i$ are the $i$-th columns of the corresponding SVD matrices and $\sigma_i$ are the ordered singular values. The particular rank-$r$ matrix that best approximates $\mathbf{R}$ is given by
% \begin{equation}
%     \mathbf{R^{r}}=\sum_{i=1}^{r}\sigma_i\mathbf{u}_i \mathbf{v}_i^{T}.
% \end{equation}

% Another interpretation of the above SVD approximation can be made in terms of Principal Component Analysis (PCA). In fact, the matrix $\mathbf{R}$ can be thought of as a matrix containing $N$ samples of $L$ variables. The covariance matrix is proportional to \begin{equation}
%     \mathbf{C} = \mathbf{R}^{H}\mathbf{R} = \mathbf{V}\mathbf{\Sigma}\mathbf{U}^H\mathbf{U}\mathbf{\Sigma}\mathbf{V}^H = \mathbf{V}\mathbf{\Sigma}^{2}\mathbf{V}^H,
% \end{equation} 
% meaning that $\mathbf{V}$ contains the principal directions of the data. The projections on the principal axes, known as principal components, are given by $\mathbf{R}\mathbf{V} = \mathbf{U}\mathbf{\Sigma}\mathbf{V}^H\mathbf{V}=\mathbf{U}\mathbf{\Sigma}$. The multiplication of the first $r$ principal components by the corresponding principal axes yields the rank $r$ approximation $\mathbf{R^{r}} = \sum_{i=1}^{r}\sigma_i\mathbf{u}_i \mathbf{v}_i^{T}$.

\subsection{Separation of GCC Components}

Let us assume that the sub-band GCC matrix can be expressed by two separable components: a target-path delay component and a noise component. Moreover, as it will be later discussed in Section \ref{sec:GCC_extraction}, we assume that the direct-path component is the one contributing to the greatest singular value, while the noise component is obtained by the addition of the rest of separable matrices, i.e.

\begin{eqnarray}
    &&\mathbf{R}^{\mathrm{target}} = \mathbf{R^1} = \sigma_1 \mathbf{u}_1 \mathbf{v}_1^{T}, \\
     &&\mathbf{R}^{\mathrm{noise}} = \sum_{i=2}^{L}\sigma_i \mathbf{u}_i \mathbf{v}_i^{T}.
\end{eqnarray}

\begin{figure}[!t]
\includegraphics[width=\columnwidth]{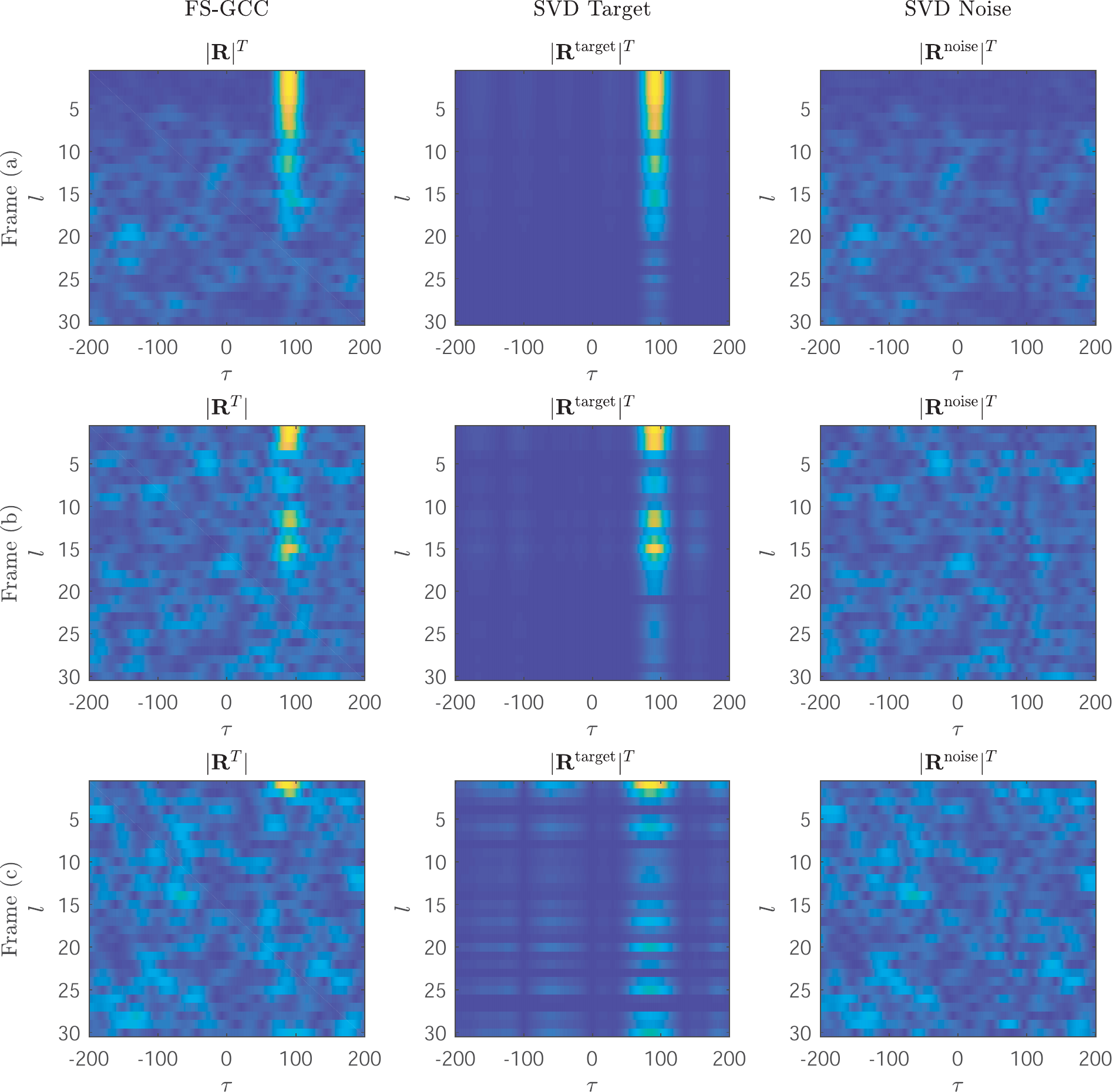}
\caption{\textcolor{black}{Extracted SVD components from three different speech frames ($\textrm{SNR} = 25$ dB) with true TDOA $\tau_0 = 92$. Left column: FS-GCC matrix. Middle column: target time-delay component. Right column: noise component.}}
\label{fig:GCCspeech_svd}
\end{figure}

An example decomposition applied over the three speech frames discussed in the previous section is shown in Fig.~~\ref{fig:GCCspeech_svd}. Note that the true delay is substantially enhanced by the target delay component of the SVD. The third speech frame (row 3) is the one having the noisiest FS-GCC, due to its reduced SNR.

\subsection{Target GCC Extraction}
\label{sec:GCC_extraction}

The singular values and the left singular vectors correspond, respectively, to the square-root of the eigenvalues and the orthonormal eigenvectors 
of $\mathbf{R}\mathbf{R}^{H}$. In the ideal case, taking into account Eq.~(\ref{eq:mod_ideal}):
\begin{equation}
    \tilde{\mathbf{R}}\tilde{\mathbf{R}}^{H} = \bm{\phi}_0\mathbf{e}^{H} \mathbf{e}\bm{\phi}_0^{H} = L\bm{\phi}_0\bm{\phi}_0^{H}
    \label{eq:hermitian}
\end{equation}
is singular, with $(L-1)$ zero eigenvalues and one non-zero eigenvalue $\lambda = L \bm{\phi}_0^{H}\bm{\phi}_0 = L\| \bm{\phi}\|^2$ associated to the eigenvector  $\bm{\phi}_0$. This means that such eigenvector should contain the information of the TDOA-shifted window of Eq.~(\ref{eq:Phi_vector}). 

In the general noisy case, the left singular vectors will be eigenvectors of
%\begin{eqnarray}
%\footnotesize
%     \mathbf{R}_{12}\mathbf{R}_{12}^{H} &=& \left(\tilde{\mathbf{R}}_{12}\mathbf{G} + \mathbf{N}(\mathbf{I}-\mathbf{G})\right) \left(\tilde{\mathbf{R}}_{12}\mathbf{G} + \mathbf{N}(\mathbf{I}-\mathbf{G})\right)^{H} \nonumber \\ 
%    &=&\left(\sum_{l=0}^{L-1} \alpha_l^2 \right)\bm{\phi}_{12}\bm{\phi}_{12}^{H} + \sum_{l=0}^{L-1}(1 - \alpha_l)^2 \mathbf{n}_l \mathbf{n}_l^{H} \nonumber \\
%    &+& \sum_{l=0}^{L-1}(\alpha_l - \alpha_l^2)\left(\mathbf{n}_l\tilde{\mathbf{r}}_l^{H} + \tilde{\mathbf{r}}_l\mathbf{n}_l^{H}\right).
%\end{eqnarray}
\begin{eqnarray}
\footnotesize
     \mathbf{R}\mathbf{R}^{H} &=& \left(\tilde{\mathbf{R}}\mathbf{G} + \mathbf{N}(\mathbf{I}-\mathbf{G})\right) \left(\tilde{\mathbf{R}}\mathbf{G} + \mathbf{N}(\mathbf{I}-\mathbf{G})\right)^{H} \nonumber \\ 
    &=&\left(\sum_{l=0}^{L-1} \alpha_l^2 \right)\bm{\phi}_0\bm{\phi}_0^{H} + \sum_{l=0}^{L-1}(1 - \alpha_l)^2 \mathbf{n}_l \mathbf{n}_l^{H} \nonumber \\
    &+& \sum_{l=0}^{L-1}(\alpha_l - \alpha_l^2)e^{j\omega_l\tau_0}\mathbf{n}_l\bm{\phi}_0^{H} + \nonumber \\ &+& \sum_{l=0}^{L-1}(\alpha_l - \alpha_l^2)e^{-j\omega_l\tau_0}\bm{\phi}_0\mathbf{n}_l^{H}.
    \label{eq:cov}
\end{eqnarray}

It is readily seen that when $\mathbf{G} = \mathbf{I}$, i.e. when no noise is present in any of the sub-bands, the result is the same as that of Eq.~(\ref{eq:hermitian}), given by the first term of Eq.~(\ref{eq:cov}). The second term, which depends on the noise components of the GCC, is a rank $L$ hermitian matrix with $L$ non-zero real eigenvalues. The third term is a rank-one matrix with a non-zero eigenvalue associated to a noise-dependent complex eigenvector. Finally, the fourth term is another rank-one matrix, with one non-zero eigenvalue related to the eigenvector $\bm{\phi}_0$. 

%It can be also observed that, for a signal having $L_g$ noiseless sub-bands (with $\alpha_l = 1$) and $L-L_g$ noisy bands ($\alpha_l < 1$), the rank reduces to $L-L_g+1$.

As a simplified example, a matrix $\mathbf{G}$ with $L_g$ ones and $(L-L_g)$ zeros in its diagonal, i.e. either with perfect sub-bands or completely noisy sub-bands, will only contain the first (target) rank-one term and the second (noise) term with rank $(L~-~L_g)$. The eigenvalue corresponding to the target term will be $L_g\| \bm{\phi}\|^2$, while the sum of the eigenvalues of the second term will be $(L-L_g)\| \bm{\phi}\|^2$. In most practical cases, the greatest singular value is expected to be dominated by the target rank-one component, which justifies the use of a rank one approximation in terms of $\mathbf{u}_{1}$.

Taking the above discussion into account, under high SNR conditions, the first left singular vector should be dominated by the first term of Eq.~(\ref{eq:cov}) corresponding to the eigenvector $\bm{\phi}_0$. Since such eigenvector should be ideally real, we take the real part of $\mathbf{u}_{1}$, $\Re\left\{\mathbf{u}_{1} \right\}$:
\begin{equation}
     \hat{\bm{\phi}}_0 [n] = \Re\left\{\mathbf{u}_{1} \right\} \mathrm{sign}\left(\Re\left\{u_{1}[\gamma]\right\} \right),
     \label{eq:estimation1}
\end{equation}
where $\Re\left\{u_{1}[\gamma]\right\}$ is the element of $\Re\left\{\mathbf{u}_1\right\}$ having the maximum absolute value, i.e.
\begin{equation}
    \gamma = \arg\max_{n}|\Re\left\{u_{1}[n]\right\}|.
\end{equation}

Note that this last equation and the $\mathrm{sign}(\cdot)$ function in Eq.~(\ref{eq:estimation1}) are used as a criterion for solving the singular vector sign ambiguity (both $\mathbf{u}_1$ and $-\mathbf{u}_1$ are eigenvectors of $\mathbf{R}\mathbf{R}^{H}$), forcing a positive peak in $\hat{\bm{\phi}}_0$. The estimated TDOA is given by the position of such peak:
\begin{equation}
    \hat{\tau}_0 = \arg\max_{n}\hat{\phi}_0[n].
\end{equation}

\textcolor{black}{The fourth column of Fig.~\ref{fig:Recovered}} shows the recovered GCCs from the target SVD (SVD FS-GCC), $\hat{\bm{\phi}}_0$, for the three speech frames shown in the previous examples. \textcolor{black}{For comparison purposes, the results for conventional GCC-PHAT are shown in the third column.}

%\begin{figure}[!t]
%\includegraphics[width=\columnwidth]{GCCsvd1.pdf}
%\caption{Sub-band GCCs for different frames of a male speech signal.}
%\label{fig:GCCsvd}
%\end{figure}

\subsection{Weighted Low-Rank Approximation}

The Frobenius norm weights uniformly all the elements of the approximation error $(\mathbf{R}-\hat{\mathbf{R}} )$. However, it can be very useful in practice to assign particular weights to the different sub-band GCCs, based on some defined confidence measure.  Therefore, weighted low-rank approximations are of interest here, where the problem is now to solve \cite{Srebro2003}
\begin{equation}
    \min_{\hat{\mathbf{R}}} \quad \left\| \left( \mathbf{R}-\hat{\mathbf{R}}\right)\odot \mathbf{W} \right\|_F, \quad \textrm{subject to} \quad \mathrm{rank}(\hat{\mathbf{R}})\leq r,
    \label{eq:wsvd}
\end{equation}
where $\mathbf{W} \in \mathbb{R}_{+}^{N \times L}$ is a weight matrix with non-negative weights and $\odot$ denotes the Hadamard product operator. The above equation is equivalent to solving a weighted Frobenius norm, where a given weight is assigned to each entry of the matrix $\mathbf{R}$, i.e.
%\begin{equation}
%    \left\| \left( \mathbf{R}_{12}-\hat{\mathbf{R}}_{12}\right)\odot \mathbf{W} \right\|_F = \sum_{n=1}^{N}\sum_{l=1}^{L}w_{n,l}^{2}\left(R_{n,l}-\hat{R}_{n,l} \right)^2,
%\end{equation}
\begin{equation}
\footnotesize
    \left\| \left( \mathbf{R}-\hat{\mathbf{R}}\right)\odot \mathbf{W} \right\|_F = \sum_{n=1}^{N}\sum_{l=1}^{L}w_{n,l}^{2}\left(\mathbf{R}[n,l]-\hat{\mathbf{R}}[n,l] \right)^2,
\end{equation}
where $w_{n,l}$ are the weight entries of the matrix $\mathbf{W}$. To address the problem, it is useful to consider the decomposition $\mathbf{R}~=~ \mathbf{A}\mathbf{B}^{H}$, where $\mathbf{A} \in \mathbb{C}^{N\times r}$ and $\mathbf{B} \in \mathbb{C}^{L\times r}$. \textcolor{black}{Any pair of matrices $(\mathbf{A},\mathbf{B})$ providing such a decomposition of a rank-$r$ matrix could be a potential solution, so the problem can be formulated as an unconstrained minimization over pairs of matrices $(\mathbf{A},\mathbf{B})$ \cite{Srebro2003}:}
\begin{equation}
    \min_{\mathbf{A},\mathbf{B}} \quad \left\| \left( \mathbf{R}-\mathbf{A}\mathbf{B}^{H}\right)\odot \mathbf{W} \right\|_F.
\end{equation}

% Since any rank-$r$
% matrix can be decomposed in such a way, and any
% pair of such matrices yields a rank-$r$ matrix, we can
% think of the problem as an unconstrained minimization
% problem over pairs of matrices $(\mathbf{A}, \mathbf{B})$ with the
% minimization objective:
% \begin{equation}
%     \min_{\mathbf{A},\mathbf{B}} \quad \left\| \left( \mathbf{R}-\mathbf{A}\mathbf{B}^{H}\right)\odot \mathbf{W} \right\|_F.
% \end{equation}
% Such decomposition is not unique. For any invertible $\mathbf{M}~\in~ \mathbb{C}^{r\times r}$, the pair $(\mathbf{AM},\,\mathbf{B}\mathbf{M}^{-1})$ provides a factorization equivalent to $\mathbf{A}\mathbf{B}^{H}$. In particular, any (non-degenerate) solution $(\mathbf{A},\mathbf{B})$ can be orthogonalized to a (non-unique) equivalent orthogonal solution $\bar{\mathbf{A}} = \mathbf{AM}$, $\bar{\mathbf{B}} = \mathbf{B}\mathbf{M}^{-1}$ such
% that $\bar{\mathbf{B}}^{H}\bar{\mathbf{B}} = \mathbf{I}$ and $\bar{\mathbf{A}}^{H}\bar{\mathbf{A}}$ is a diagonal matrix. 
% As discussed in the previous section, the solution in the unweighted case would be obtained in terms of the truncated SVD. 

In our particular application, the weight matrix can be restricted to follow the specific structure:
\begin{equation}
    \mathbf{W} = \begin{bmatrix} 
        w_{0} & w_{1} & \dots & w_{L-1} \\
        w_{0} & w_{1} & \dots & w_{L-1} \\
        \vdots & \vdots & \vdots & \vdots \\
        w_{0} & w_{1} & \dots & w_{L-1}
    \end{bmatrix} = \mathbf{1}\mathbf{w}^{T},
\end{equation}
where $\mathbf{1}$ is a column vector of length $N$ and $\mathbf{w}~ =~[w_0,\,w_1,\dots,\,w_{L-1}]^{T}$ is a vector containing confidence weights $w_l\in [0,1]$ assigned to each of the $L$ sub-bands. Under such case, due to the structure of $\mathbf{W}$, the problem of Eq.~(\ref{eq:wsvd}) can be rewritten as the factorization of a modified matrix $\mathbf{R}_{w}$ \cite{Auiar99}:
\begin{equation}
    \min_{\mathbf{A},\mathbf{B}_{w}} \quad \left\| \left( \mathbf{R}_{w}-\mathbf{A}\mathbf{B}_w^{H}\right) \right\|_F,
\end{equation}
where
\begin{eqnarray}
    \mathbf{R}_{w}&=&\mathbf{R}\tilde{\mathbf{W}},\\
    \mathbf{B}_w &=& \tilde{\mathbf{W}}\mathbf{B},\\
    \tilde{\mathbf{W}} &=& \mathrm{diag}(w_1,\,w_2,\dots,w_L).
\end{eqnarray}

By applying SVD factorization to $\mathbf{R}_{w}$ and conveniently truncating to the largest $r$ singular values, the estimates of $\mathbf{A}$ and $\mathbf{B}_w$ are obtained as $\hat{\mathbf{A}} = \mathbf{U}\hat{\mathbf{\Sigma}}^{1/2}$ and $\hat{\mathbf{B}}_w = \mathbf{V}\hat{\mathbf{\Sigma}}^{1/2}$, where $\hat{\mathbf{\Sigma}}$ is the truncated singular value matrix. The rank-$r$ approximation is therefore recovered as:
\begin{equation}
    \mathbf{R^r} = \hat{\mathbf{A}}\hat{\mathbf{B}}^{H} = \hat{\mathbf{A}}(\hat{\mathbf{W}}^{-1}\hat{\mathbf{B}}_w)^{H}. 
\end{equation}

Taking again into consideration that the noiseless sub-band GCC matrix is rank one, the target and noise components are extracted as:
\begin{eqnarray}
    \mathbf{R}^{\mathrm{target}} &=& 
    \hat{\mathbf{a}}_{1}\hat{\mathbf{b}}_1^{H}, \\
     \mathbf{R}^{\mathrm{noise}} &=& \sum_{i=2}^{L} \hat{\mathbf{a}}_{i}\hat{\mathbf{b}}_i^{H},
\end{eqnarray}
where $\hat{\mathbf{a}}_i$ and $\hat{\mathbf{b}}_i$ are, respectively, the columns of $\hat{\mathbf{A}}$ and $\hat{\mathbf{B}}$. Similarly to the SVD case, the recovered GCC is obtained as
\begin{equation}
     \hat{\bm{\phi}}_0[n] =  \Re\left\{\hat{\mathbf{a}}_{1} \right\} \mathrm{sign}\left(\Re\left\{\hat{a}_{1}[\gamma]\right\} \right),
     \label{eq:estimation}
\end{equation}
where $\Re\left\{\hat{a}_{1}[\gamma]\right\}$ is the element of $\Re\left\{\hat{\mathbf{a}}_{1}\right\}$ having the maximum absolute value, i.e.
\begin{equation}
    \gamma = \arg\max_{n}|\Re\left\{\hat{a}_{1}[n]\right\}|.
\end{equation}

\begin{figure*}[!t]
\includegraphics[width=\textwidth]{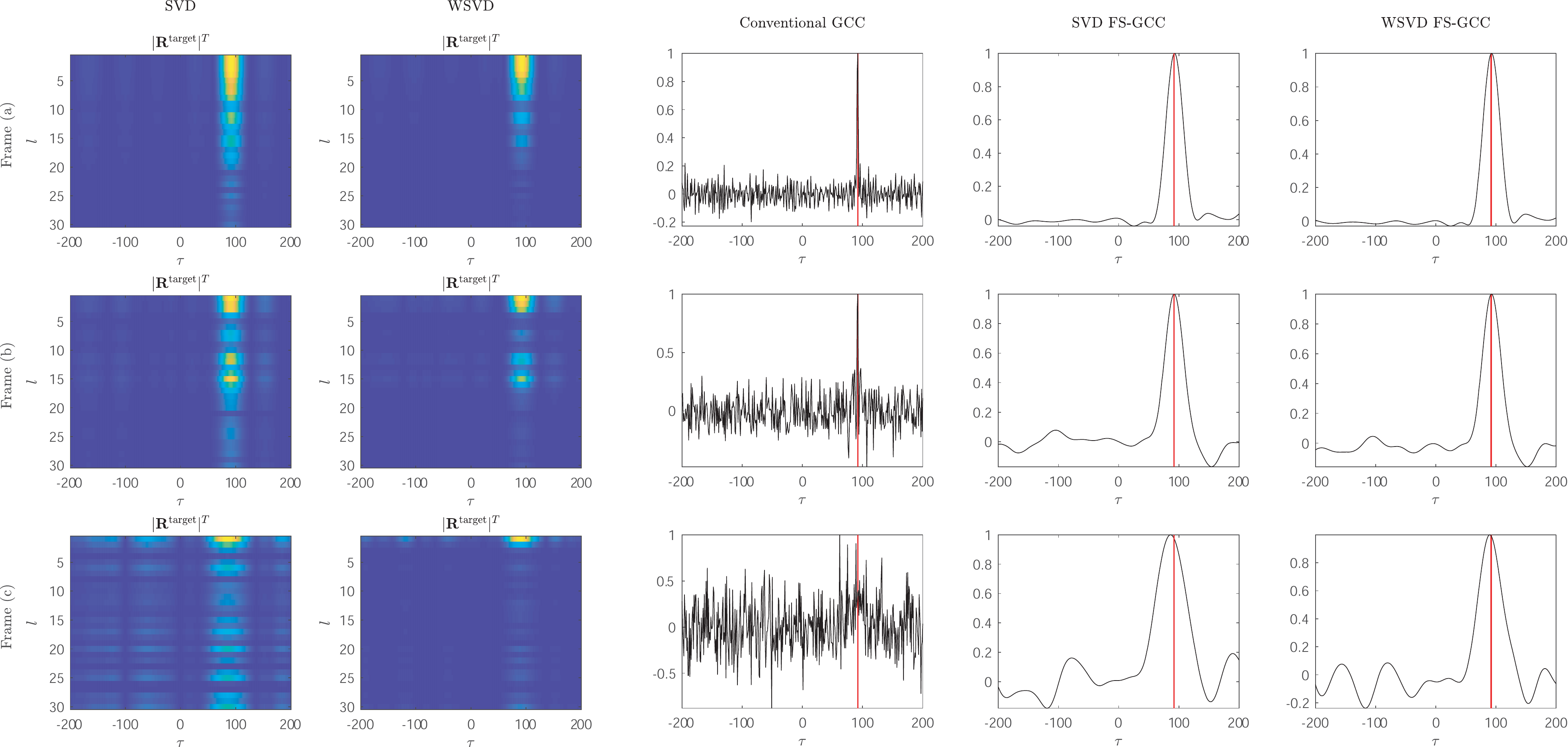}
\caption{\textcolor{black}{Results for three speech frames using SVD FS-GCC and WSVD-FSGCC. The two first columns show the extracted target components using SVD and WSVD. The rest of columns show the recovered GCCs and the true TDOA (red line, $\tau_0 = 92$ samples). Note that the conventional GCC (third column) fails in speech frame (c).}}
\label{fig:Recovered}
\end{figure*}

\textcolor{black}{The target components resulting from the considered speech examples are shown in the second column of Fig.~\ref{fig:Recovered}, while the recovered GCCs from the weighted SVD (WSVD FS-GCC) are shown in the last column.\footnote{Code available at https://github.com/spatUV/fs-gcc}}

%\begin{figure}[!t]
%\includegraphics[width=\columnwidth]{GCCwsvd1.pdf}
%\caption{Sub-band GCCs for different frames of a male speech signal.}
%\label{fig:GCCwsvd}
%\end{figure}

\subsubsection{Weighting} 

Note that the weights $w_l$ can be directly related to the $\alpha_l$ coefficients of the model in Eq.~(\ref{eq:model_noisy}). Weights having a value close to 1 must reflect that the corresponding band provides almost perfect delay information. In contrast, weights close to zero should reflect that the frequency band is dominated by noise and its information should be discarded. Similarly, perfect sub-bands are expected to have a magnitude GCC which is only dependent on the displaced window response $\bm{\phi}_0$, as expressed by Eq.~(\ref{eq:perfect_column}). Therefore, the average of the magnitude of a perfect sub-band GCC ($\alpha_l=1$), regardless the true TDOA, will be:
\begin{equation}
   \overline{|\mathbf{r}_l|}\bigr|_{\alpha=1} = \frac{1}{N}\sum_{n=0}^{N-1}\bigl|\bm{\phi}[n]\bigr|.
\end{equation}
In contrast, GCCs from completely noisy sub-bands ($\alpha_l=0$) are given by noise realizations $\mathbf{n}_l$. The magnitude of a noisy band, which follows a Rayleigh distribution, will tend to its mean value:
\begin{equation}
   \overline{|\mathbf{r}_l|}\bigr|_{\alpha=0} = \sqrt{\frac{1}{2N}\|\bm{\phi}\|^2}\sqrt{\frac{\pi}{2}}.
\end{equation}

Taking the above two extreme cases into account, the proposed weights are:
\begin{equation}
    \centering
w_l = \begin{cases}
    g_l, & g_l \geq 0\\
    0, & g_l< 0,
  \end{cases}
  \end{equation}
where
\begin{equation}
    g_l = \frac{\left(\overline{|\mathbf{r}_l|}\bigr|_{\alpha=0}-\frac{1}{N}\sum_{n=0}^{N-1}\left|r_l[n]\right|\right)}{\left(\overline{|\mathbf{r}_l|}\bigr|_{\alpha=0}-\overline{|\mathbf{r}_l|}\bigr|_{\alpha=1}\right)}. 
\end{equation}

Therefore, the weights are expected to vary between 1 (perfect sub-bands) and 0 (completely noisy sub-bands). \textcolor{black}{Note that the above equations were derived by assuming that ``noisy" sub-bands only comprise frequency bins where either the SNR is very low or there is no signal at all. These bands are not expected to incorporate early echoes or reverberation effects, at least, as long as the frame size is longer than the length of the acoustic channel. Under this assumption, the cross-power spectrum phase will be uniformly distributed in the range $[-\pi,\pi]$, and so the sub-band GCC coefficients wil correspond to a filtered complex Gaussian process \cite{Brillingerbook} with Rayleigh-distributed magnitude.}

\section{Experiments}
\label{sec:experiments}

This section describes the experiments conducted to show the advantages of the proposed FS-GCC method in terms of TDE performance. The criteria used to assess such performance and the complete experimental set-up are next described.

\subsection{Performance Criteria}

We classify a time delay estimate as either an anomaly or a nonanomaly according to its absolute error $e_i = | \tau - \hat{\tau}_i|$,  where $\tau$ is the true time delay and $\hat{\tau}_i$ is the $i$-th time delay estimate. If $e_i > T_c/2$, the estimate is assumed to be anomalous, where $T_c$ is the signal correlation time \cite{Champagne1996}. For our particular source signal (speech), $T_c$ was computed as the width of the main lobe of its autocorrelation function (taken between the -3 dB points), which is equal to 24 samples. The TDE performance is evaluated in terms of the percentage of anomalous estimates over the total estimates ($P_{\hat{\tau}}$), \textcolor{black}{the average GCC first-to-second peak ratio (FSPR)}, the mean and the standard deviation of the absolute error (all for the subset of nonanomalous estimates: \textcolor{black}{$\mathrm{FSPR}_{\hat{\tau},\mathrm{na}}$}, $\mathrm{MAE}_{\hat{\tau},\mathrm{na}}$,  $\mathrm{SDAE}_{\hat{\tau},\mathrm{na}}$). These measures are defined as
%the absolute error for all estimates ($\mathrm{MAE}_{\hat{\tau}}$, $\mathrm{SDAE}_{\hat{\tau}}$) and 

\begin{eqnarray}
    P_{\hat{\tau}} &=& \frac{N_\mathrm{a}}{N_\mathrm{T}}, \\
    %\mathrm{MAE}_{\hat{\tau}} &=& \frac{1}{N_\mathrm{T}}\sum_{i}e_i, \\
    %\mathrm{SDAE}_{\hat{\tau}} &=& \sqrt{\frac{1}{N_\mathrm{T}}\sum_{i}\left(e_i-\mathrm{MAE}_{\hat{\tau}}\right)^2},\\
    \mathrm{MAE}_{\hat{\tau},\mathrm{na}} &=& \frac{1}{N_{\mathrm{na}}}\sum_{i\in \chi_{\mathrm{na}}}e_i, \\
    \mathrm{SDAE}_{\hat{\tau},\mathrm{na}} &=& \sqrt{\frac{1}{N_{\mathrm{na}}}\sum_{i\in \chi_{\mathrm{na}}}\left(e_i-\mathrm{MAE}_{\hat{\tau},\mathrm{na}}\right)^2},
\end{eqnarray}
where $N_\mathrm{T}$ denotes the total number of estimates, $N_\mathrm{a}$ is the number of estimates that are identified as anomalies, $N_\mathrm{na}$ is
the number of nonanomalous estimates, and $\chi_{\mathrm{na}}$ represents the
subset of nonanomalous estimates. \textcolor{black}{The FSPR is defined as the average gain (over the subset of nonanomalous estimates) of the maximum GCC peak with respect to the second larger peak, i.e.
\begin{equation}
    \mathrm{FSPR}_{\hat{\tau},\mathrm{na}} = \frac{1}{N_{\mathrm{na}}} \sum_{i\in \chi_{\mathrm{na}}} 20\log_{10}\left(\frac{\hat{\bm{\phi}}_0^{(i)}[\hat{\tau}_0]}{\hat{\bm{\phi}}_0^{(i)}[\hat{\tau}_1]}\right),
\end{equation}
where $\hat{\tau}_1$ denotes the lag of the second larger peak in the GCC.}

\begin{figure*}[!t]
\includegraphics[width=\textwidth]{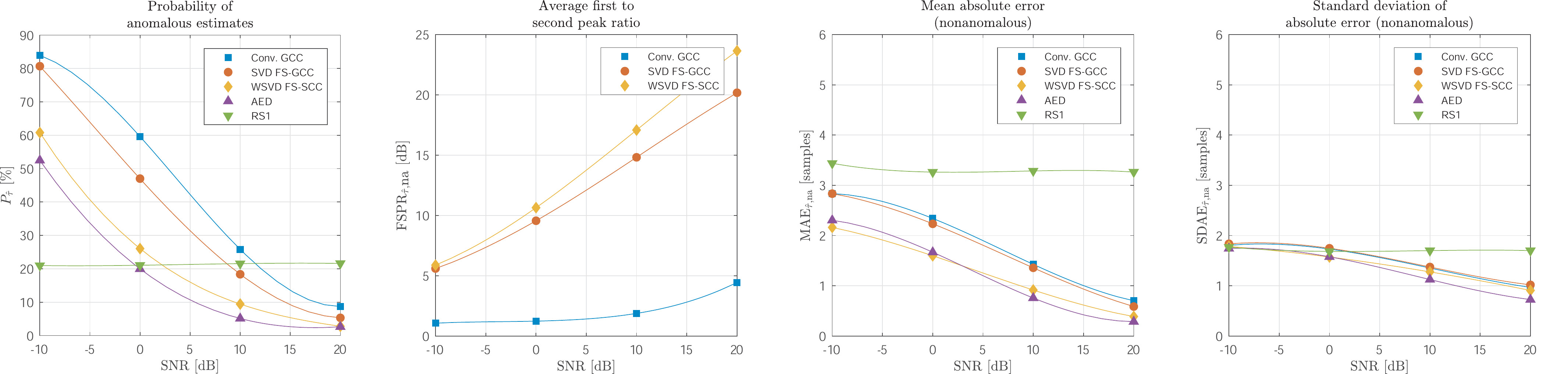}
\caption{Performance evaluation in anechoic scenario ($r_i = 0$). The fitting curve is a third order polynomial.}
\label{fig:perf_ane}
\end{figure*}

\begin{figure*}[!t]
\includegraphics[width=\textwidth]{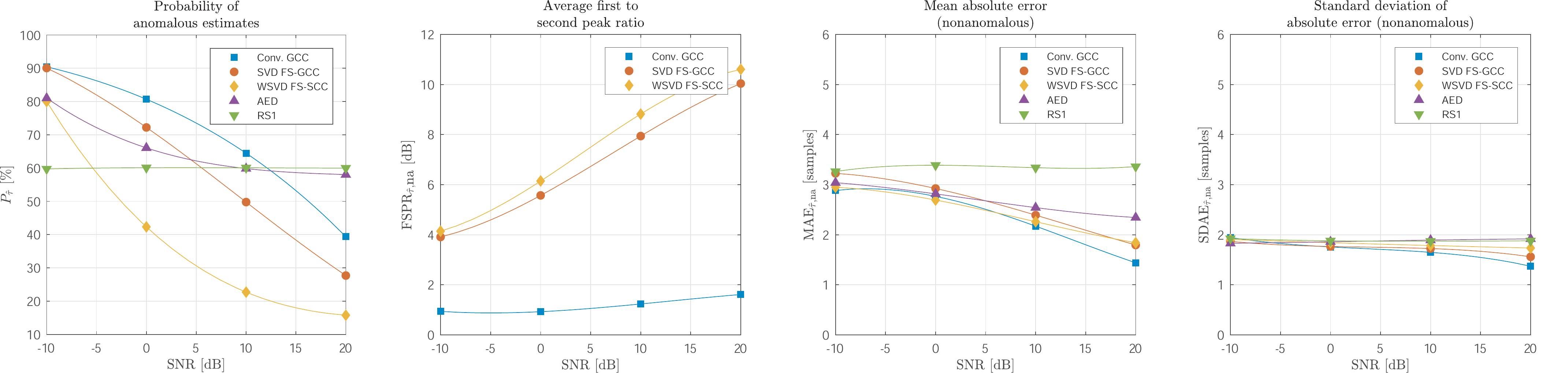}
\caption{Performance evaluation in reverberant scenario ($r_i = 0.8$, \textcolor{black}{$T_{60} \approx 0.3$ s}). The fitting curve is a third order polynomial.}
\label{fig:perf_rev}
\end{figure*}

\subsection{Simulation Set-up and Algorithm Parameters}

We consider a rectangular room simulated by the image-source method in a single source scenario \cite{Allen1979}. Synthetic impulse responses were generated for a pair of sensors separated 0.5 meters considering 10 random array positions and orientations within the room, as well as 10 random source locations for each microphone configuration. The simulations were repeated for each reverberant condition. The following parameters were used:
\begin{itemize}
    \item Room dimensions: 6 $\times$ 7 $\times$ 3 meters ($x$ $\times$ $y$ $\times$ $z$).
    \item Uniform reflection coefficients: $r_i \in \{0.0,\,0.8\}$. %($i=1,\dots,5$) varying between 0 and 1.
    \item Source positions: 10 random positions on the plane ($x$, $y$, $z=1.25$).
    \item Microphone positions: two-microphone array with inter-sensor spacing 0.5 m, with 10 random positions and orientations on the $x-y$ plane ($z=1.25$).
    \item SNR: Varying between -10 dB and 20 dB. For each array set-up and source position, 10 different noise realizations were generated for each SNR condition. To control the SNR, mutually independent white Gaussian noise was properly scaled and added to each microphone signal.
    \item Source signal: A male speech signal of 2 seconds duration, digitized with 16-bit resolution at 44.1 kHz. The signal was processed to eliminate non-activity segments.
\end{itemize}

The synthetic microphone signals obtained by convolving the source signal with the generated impulse responses were processed by the different methods in the Short-Time Fourier Transform (STFT) domain. We used a frame length of 2048 samples and Hann windowing with 75\% overlap, leading to 177 frames per audio example. The total number of estimates used to evaluate each method at each SNR and reverberant condition is therefore $N_{\mathrm{T}} = 10 \times 10 \times 10 \times 177 = 177,000$.

The parameters used for evauating the FS-GCC approach were $B=128$ (spectral window length) and $M=32$ (hop) frequency bins. Results for a blind channel identification method, Adaptive Eigenvalue Decomposition (AED) \cite{BenestyAED}, and \textcolor{black}{a method based on ratios of acoustical transfer functions (RTFs) (the algorithm denoted as RS1 in \cite{dvorkind2005time}) are also given for comparison purposes.} The AED method was configured to use rectangular windows of the same size (2048), a filter length of 512 samples and an adaptation step of $\mu=0.003$. \textcolor{black}{The RS1 method was configured as in a tracking scenario, considering as frames rectangular windows of 2048 samples with 75\% overlap. The power spectral density (PSD) estimation needed for RS1 was performed using the Welch~\cite{welch1967use} method with Hann windows of 256 samples and 50\% overlap, resulting in 15 periodograms for each PSD estimate. The Recursive Least Square (RLS) method, applied in RS1, was configured using a forgetting factor of $\alpha=0.8222$, following~\cite{dvorkind2005time}}. Since \textcolor{black}{AED and RS1 are not GCC-based methods, the FSPR is not provided for such algorithms}.

\subsection{TDE Results}

The results for the anechoic condition with a varying SNR are shown in  Fig.~\ref{fig:perf_ane}. The percentage of anomalous estimates is clearly reduced for all the methods with respect to the conventional GCC-PHAT, with significant improvements achieved by WSVD FS-GCC, \textcolor{black}{RS1} and, especially, AED. \textcolor{black}{At the lowest SNR, RS1 is the most robust method, but its performance seems to be bounded for higher SNRs. Note that the different behavior of RS1 may be due to the aforementioned tracking scenario configuration.} AED and WSVD FS-GCC follow a similar behavior. The biggest improvement is observed for SNR $= 0$ dB, where the difference between WSVD FS-GCC and conventional GCC is close to 35 percentage points (40 in the case of AED \textcolor{black}{and RS1}). The \textcolor{black}{FSPR} is significantly better for the FS-GCC methods than for the conventional GCC, especially at higher SNRs. All the methods outperform as well the conventional GCC-PHAT in terms of mean and standard deviation of time-delay errors, following a similar behavior (\textcolor{black}{except RS1, which has a higher average error}). This is an interesting result, since it means that having a lower temporal resolution due to the windowing effect does not affect negatively the TDE accuracy of FS-GCC.

The results for the reverberant case $r_i = 0.8$ \textcolor{black}{(reverberation time, $T_{60} \approx 0.3$ s)} are shown in Fig.~\ref{fig:perf_rev}. The percentage of anomalous estimates are now worse for all the methods, but in this case, \textcolor{black}{both AED and RS1} are significantly affected by reverberation. In contrast, WSVD FS-GCC still shows very good robustness, with the most significant difference (more than 40 percentage points) at SNR $= 10$ dB. The \textcolor{black}{FSPR} is again considerably better for both FS-GCC methods, with SVD providing slightly better results than WSVD. Regarding nonanomalous TDOA errors, the FS-GCC methods provide slightly better results at low SNRs. At higher SNRs the conventional GCC-PHAT seems to be slightly more accurate, although in all cases the differences are very small (below 1 sample).

\subsection{Impact on Source Localization Performance}

\begin{figure}[!t]
\includegraphics[width=\columnwidth]{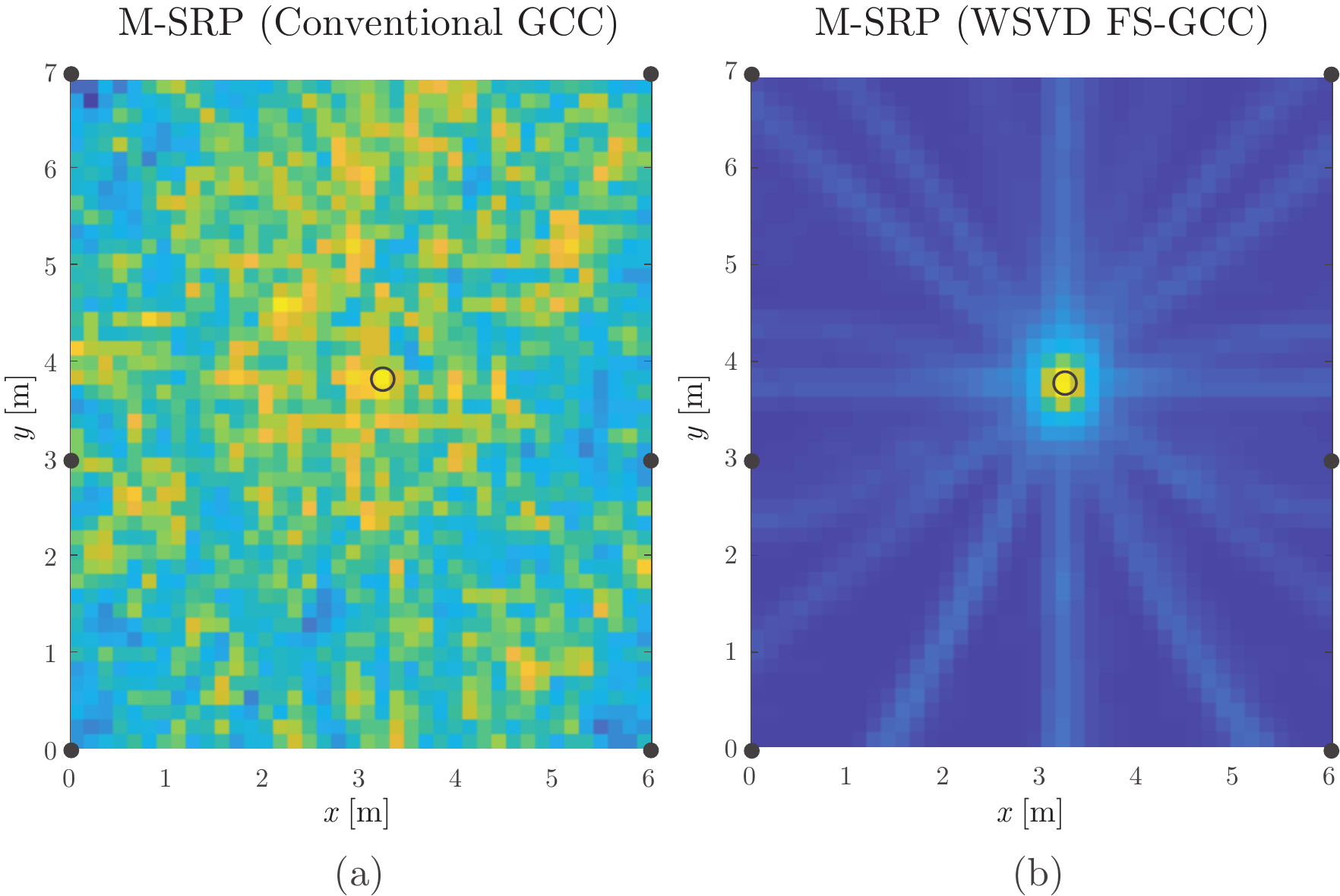}
\caption{Example M-SRP maps (SNR = 0 dB). Microphones are represented with black dots. The circle indicates the source location. (a) Using conventional GCCs. (b) Using WSVD FS-GCCs.}
\label{fig:SRP}
\end{figure}

Although accurate time delay estimates are assumed to lead to better localization results, the advantages in terms of \textcolor{black}{FSPR} are also expected to contribute to better localization in SRP-based approaches, due to the mitigation of noise in the GCCs. To support such claim, experiments were conducted considering the same acoustic conditions and source signals but with six distributed microphones placed at the walls and corners of the room. The modified SRP algorithm (M-SRP) \cite{CobosSPL2011} was applied considering a grid resolution of 0.15 m. Fig.~(\ref{fig:SRP}) shows an example of the resulting SRP maps when using conventional GCCs (a) and the proposed WSVD FS-GCCs (b) for the same signal frame (SNR$ = 0$ dB). The improvement in terms of robustness to noise is clearly observed. The results for the mean and median absolute location errors are specified in Tables~\ref{tab:loc_rho0_0} and \ref{tab:loc_rho0_8}, respectively, for anechoic and reverberant conditions. For comparison purposes, localization performance for AED is also provided by estimating the source location as the point of the grid having the lowest mean squared error considering all the available TDOAs. It can be observed that FS-GCC provides always more accurate location estimates and less anomalous detections (lower median) both in the anechoic case and in the reverberant case. Note that in the very worst case of SNR $= -10$ dB and reverberation, all the methods provide unreliable location estimates.

\begin{table}[]
\caption{Localization error [meters] for $r_i = 0$} 
\resizebox{\columnwidth}{!}{%
\begin{tabular}{@{}lcccccccc@{}}
\toprule
\multicolumn{1}{c}{\multirow{2}{*}{Method}} & \multicolumn{2}{c}{SNR = 20 dB} & \multicolumn{2}{c}{SNR = 10 dB} & \multicolumn{2}{c}{SNR = 0 dB} & \multicolumn{2}{c}{SNR = -10 dB} \\ \cmidrule(l){2-9} 
\multicolumn{1}{c}{} & Mean & Median & Mean & Median & Mean & Median & Mean & Median \\ \midrule
GCC & 0.2510 & 0.0839 & 0.6786 & 0.0844 & 1.9216 & 1.8230 & 2.7655 & 2.7102 \\
FS-GCC & 0.0716 & 0.0595 & 0.0819 & 0.0702 & 0.3685 & 0.0839 & 1.5298 & 0.4895 \\
AED & 0.9542 & 0.0702 & 0.9802 & 0.0702 & 1.3457 & 0.5862 & 1.9669 & 1.7781 \\ \bottomrule
\end{tabular}
}
\label{tab:loc_rho0_0}
\end{table}

\begin{table}[]
\caption{Localization error [meters] for $r_i = 0.8$, \textcolor{black}{$T_{60} \approx 0.3$ s}} 
\resizebox{\columnwidth}{!}{%
\begin{tabular}{@{}lcccccccc@{}}
\toprule
\multicolumn{1}{c}{\multirow{2}{*}{Method}} & \multicolumn{2}{c}{SNR = 20 dB} & \multicolumn{2}{c}{SNR = 10 dB} & \multicolumn{2}{c}{SNR = 0 dB} & \multicolumn{2}{c}{SNR = -10 dB} \\ \cmidrule(l){2-9} 
\multicolumn{1}{c}{} & Mean & Median & Mean & Median & Mean & Median & Mean & Median \\ \midrule
GCC & 1.5722 & 0.9079 & 2.4151 & 2.4198 & 2.7016 & 2.6541 & 2.7025 & 2.6055 \\
FS-GCC & 0.2825 & 0.0839 & 0.6675 & 0.0926 & 1.5757 & 0.9293 & 2.7053 & 2.5615 \\
AED & 1.8427 & 1.6497 & 1.8589 & 1.6620 & 1.9365 & 1.8054 & 2.6129 & 2.5434 \\ \bottomrule
\end{tabular}
}
\label{tab:loc_rho0_8}
\end{table}

{\color{black}
\subsection{Direction of Arrival Estimation in Real Recordings} 
In order to prove the effectiveness of the FS-GCC method also in a real environment, we tested Direction of Arrival (DoA) estimation performances over the LOCATA challenge dataset~\cite{Evers2019}. Specifically, we considered  Task 1, consisting in a single static source and a single static microphone array, from the evaluation version of the dataset, for a total of 13 different recordings and 43122 samples. Among available setups, we  chose the DICIT planar array, and considered as sub-arrays, microphones separated by a distance of $0.032$~m, since this is the closest setup to the ones considered in the rest of the paper, thus facilitating the comparison. DoA estimation is performed using the M-SRP algorithm with a uniform spherical grid of regular triangles formed by 5 icosahedron subdivisions. Voice-Active Periods (VAP) were extracted considering the Voice-Activity Detection (VAD) labels provided with the dataset.

Azimuth and elevation average errors during  VAP for Task 1 are shown in Tab.~\ref{tab:doa_est_locata}.  FS-GCC based M-SRP outperforms GCC both with respect to azimuth and elevation estimation, although only by a small amount. This behavior is expected, since the LOCATA dataset is not excessively noisy and, as shown in Fig.~\ref{fig:perf_rev}, FS-GCC and GCC TDE performances become progressively similar as the SNR increases.
Results related to elevation are significantly worse than the ones related to the azimuth, due to the fact that the DICIT array has a lower diversity on the z-axis (of the 15 microphones considered only two are positioned at a different height with the respect to the remaining ones).
Although~\cite{Evers2019} provides only mean azimuth average errors, the values obtained using both FS-GCC and GCC are comparable with the other methods submitted to the LOCATA challenge and considering Task 1. It should be also emphasized that no tracking mechanism was used, as opposed to some of the methods evaluated in LOCATA.

\begin{table}[]
\caption{Average azimuth and elevation errors [degrees] for task $1$ of the LOCATA challenge} 
%\resizebox{\columnwidth}{!}{%
\begin{center}
\begin{tabular}{@{}lcc@{}}
\toprule
\multicolumn{1}{c}{Method} & Azimuth & Elevation \\ \midrule
GCC & 2.514 & 9.333\\
FS-GCC & 2.391 & 8.536   \\ \bottomrule
\end{tabular}
\end{center}
%}
\label{tab:doa_est_locata}
\end{table}
} % end Color

\section{Conclusion}
\label{sec:conclusion}
This paper presented an improved GCC-based technique for TDE based on a sub-band analysis of the cross-power spectrum phase. The properties resulting from the so-called frequency-sliding GCC (FS-GCC) allows the recovering of denoised correlation signals by means of low-rank approximations of the FS-GCC matrix. The use of SVD and weighted SVD for obtaining both robust GCCs and accurate time delay estimates has been validated and compared to other well-known approaches, showing the relevant impact that the proposed technique can have in TDE and source localization performance.

% if have a single appendix:
%\appendix[Proof of the Zonklar Equations]
% or
%\appendix  % for no appendix heading
% do not use \section anymore after \appendix, only \section*
% is possibly needed

% use appendices with more than one appendix
% then use \section to start each appendix
% you must declare a \section before using any
% \subsection or using \label (\appendices by itself
% starts a section numbered zero.)
%

%\appendices
%\section{Proof of the First Zonklar Equation}
%Appendix one text goes here.

% you can choose not to have a title for an appendix
% if you want by leaving the argument blank
%\section{}
%Appendix two text goes here.

% use section* for acknowledgment
%\section*{Acknowledgment}
%The authors would like to thank...

% Can use something like this to put references on a page
% by themselves when using endfloat and the captionsoff option.
\ifCLASSOPTIONcaptionsoff
  \newpage
\fi

% trigger a \newpage just before the given reference
% number - used to balance the columns on the last page
% adjust value as needed - may need to be readjusted if
% the document is modified later
%\IEEEtriggeratref{8}
% The "triggered" command can be changed if desired:
%\IEEEtriggercmd{\enlargethispage{-5in}}

% references section

% can use a bibliography generated by BibTeX as a .bbl file
% BibTeX documentation can be easily obtained at:
% http://mirror.ctan.org/biblio/bibtex/contrib/doc/
% The IEEEtran BibTeX style support page is at:
% http://www.michaelshell.org/tex/ieeetran/bibtex/
%\bibliographystyle{IEEEtran}
% argument is your BibTeX string definitions and bibliography database(s)
%\bibliography{IEEEabrv,../bib/paper}
%
% <OR> manually copy in the resultant .bbl file
% set second argument of \begin to the number of references
% (used to reserve space for the reference number labels box)

\bibliography{refs} 
\bibliographystyle{ieeetr}

%\begin{thebibliography}{1}

%\bibitem{IEEEhowto:kopka}
%H.~Kopka and P.~W. Daly, \emph{A Guide to \LaTeX}, 3rd~ed.\hskip 1em plus
%  0.5em minus 0.4em\relax Harlow, England: Addison-Wesley, 1999.

%\end{thebibliography}

% biography section
% 
% If you have an EPS/PDF photo (graphicx package needed) extra braces are
% needed around the contents of the optional argument to biography to prevent
% the LaTeX parser from getting confused when it sees the complicated
% \includegraphics command within an optional argument. (You could create
% your own custom macro containing the \includegraphics command to make things
% simpler here.)
%\begin{IEEEbiography}[{\includegraphics[width=1in,height=1.25in,clip,keepaspectratio]{mshell}}]{Michael Shell}
% or if you just want to reserve a space for a photo:

\begin{IEEEbiography}[{\includegraphics[width=1in,height=1.25in,clip,keepaspectratio]{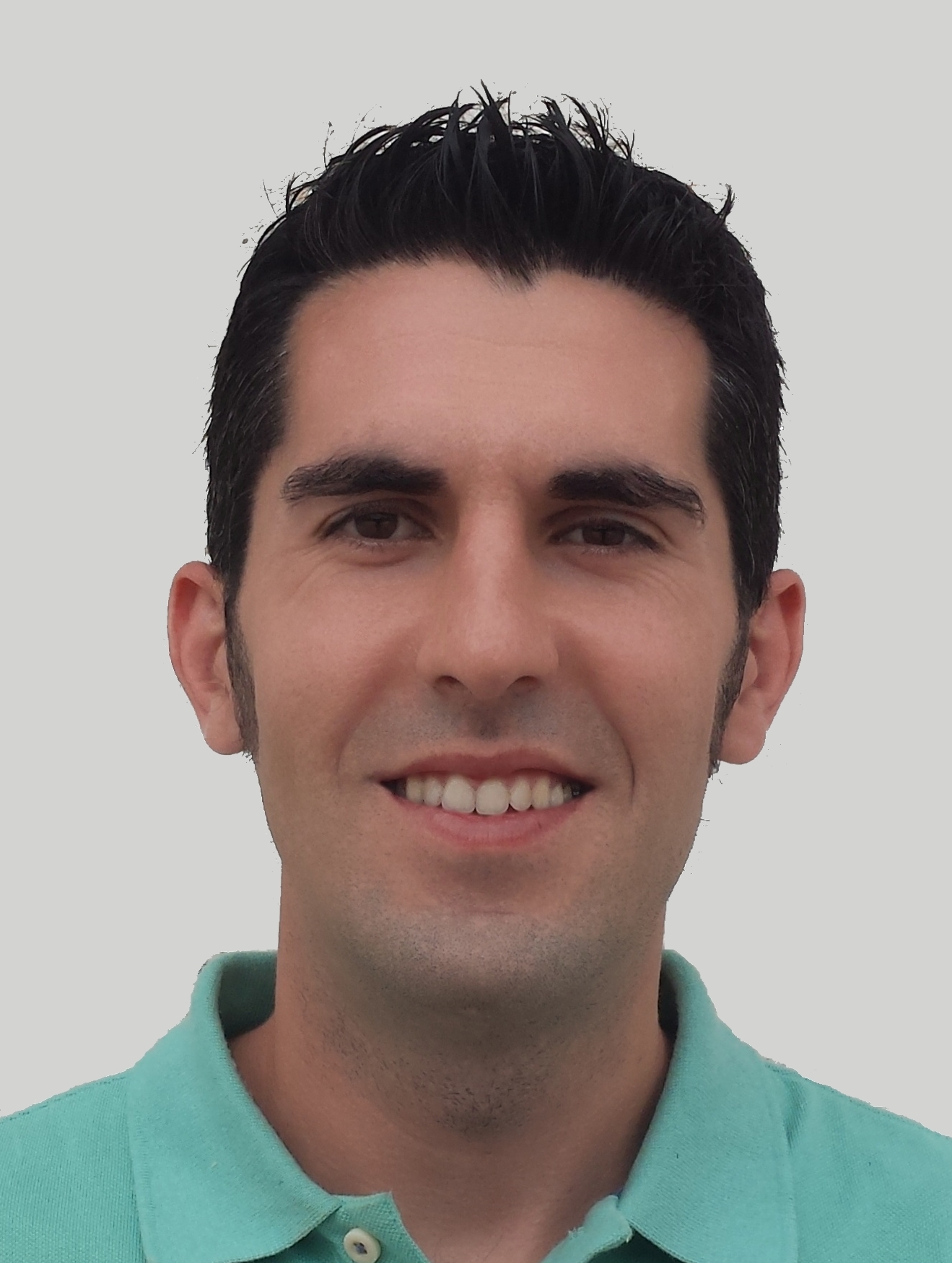}}]{Maximo Cobos} (SM’12) 
received the master’s degree in telecommunications and the Ph.D. degree in telecommunications engineering from the Universitat Polit\`ecnica de Val\`encia, Valencia, Spain, in 2009 and 2007, respectively. He completed with honors
his studies under the University Faculty Training program (FPU) and was the recipient of the Ericsson Best Ph.D. Thesis Award from the Spanish National Telecommunications Engineering Association. In 2010, he received a Campus de Excelencia postdoctoral fellowship to work at the Institute of Telecommunications and Multimedia Applications. In 2011, he joined the Universitat de Val\`encia, where he is currently an Associate Professor. His work is focused on the area of digital signal processing and machine learning for audio and multimedia applications, where he has authored/coauthored more than 100 technical papers in international journals and conferences. He is a member of the Audio Signal Processing Technical Committee of the European Acoustics Association and serves as Associate Editor for IEEE Signal Processing Letters.
\end{IEEEbiography}

\begin{IEEEbiography}[{\includegraphics[width=1in,height=1.25in,clip,keepaspectratio]{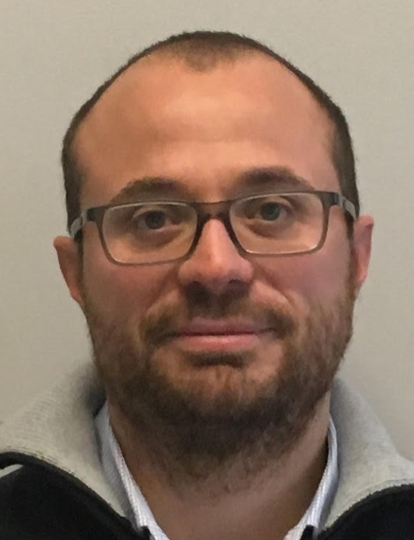}}]{Fabio Antonacci}
(M’14) was born in Bari, Italy, on July 26, 1979. He received the Laurea degree in 2004 in telecommunication engineering and the Ph.D. degree in information engineering in 2008, both from the Politecnico di Milano, Milan, Italy.
He is currently an Assistant Professor at the Politecnico di Milano. His research focuses on spacetime processing of audio signals, for both speaker and microphone arrays (source localization, acoustic scene analysis, rendering of spatial sound) and on modeling of acoustic propagation.
He is a member of the IEEE Audio and Acoustic Signal Processing Technical Committee and of the EURASIP SAT on Audio, Speech and Music Signal Processing.
\end{IEEEbiography}

\begin{IEEEbiography}[{\includegraphics[width=1in,height=1.25in,clip,keepaspectratio]{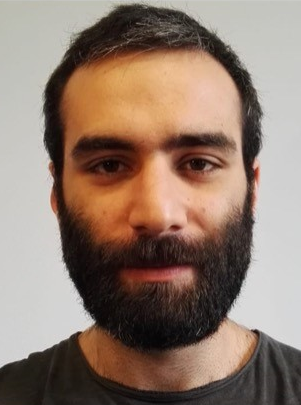}}]{Luca Comanducci}
(S’19) received the B.S. degree in Music Information Science from the University of Milan, Milan, Italy, in 2014 and  the M.S. degree in Computer Science and Engineering from Politecnico di Milano, Milan, Italy, in 2018, where he is currently working toward the Ph.D. degree in information engineering within the Dipartimento di Elettronica, Informazione and Bioingegneria. His main research interests concern the application of Deep Learning techniques to space-time audio signal processing.
\end{IEEEbiography}

\begin{IEEEbiography}[{\includegraphics[width=1in,height=1.25in,clip,keepaspectratio]{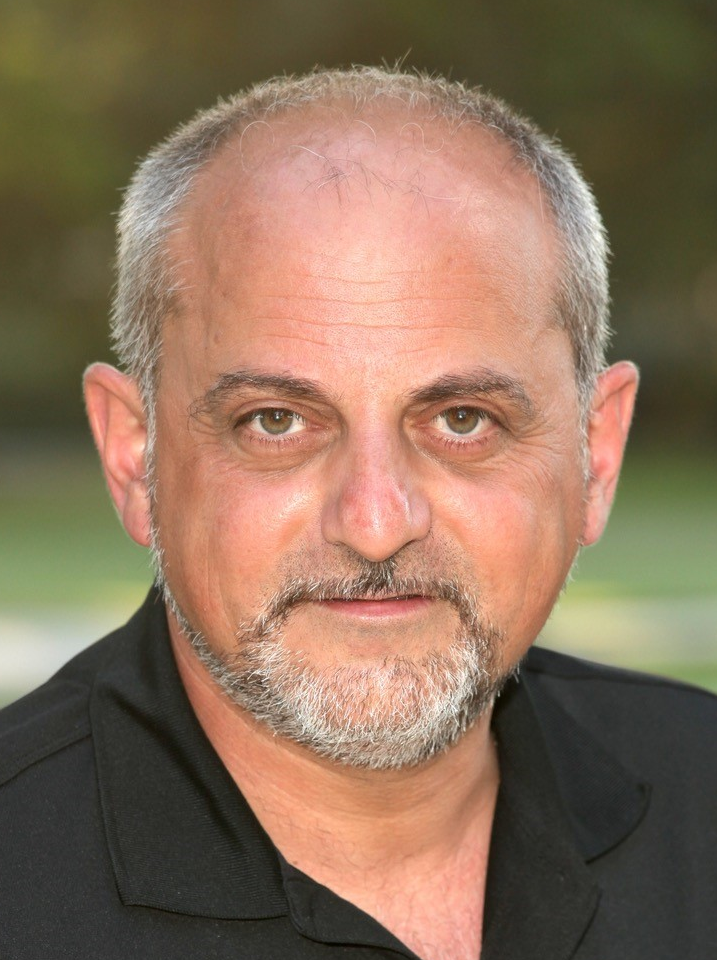}}]{Augusto Sarti}
(M’04-SM’13) received his Ph.D. in Information Engineering from the University of Padova, Italy, in 1993 (joint program with University of California, Berkeley). After a few months as Post-Doc researcher at UC Berkeley, he joined the Faculty of the Politecnico di Milano (PoliMI), Italy, where he is currently a Full Professor. In 2013, he also joined the University of California, Davis, as a Full Professor (adjunct). He is the co-founder of the Image and Sound Processing Group at PoliMI, with which he promoted/coordinated and/or contributed to numerous European projects. He currently coordinates the activities of ISPG's audio labs. He is also the founder and coordinator of PoliMI's M.Sci. program in Music and Acoustic Engineering. He coauthored over 300 scientific publications on international journals and congresses and numerous patents in the multimedia signal processing area. His research interests are in the area of multimedia signal processing, with particular focus on sound analysis, synthesis, and processing; space-time audio processing; geometrical acoustics; and music information retrieval. He has been an Active Member of the IEEE Technical Committee on Audio and Acoustics Signal Processing and of the Editorial Board of the IEEE.
\end{IEEEbiography}

% if you will not have a photo at all:
%\begin{IEEEbiographynophoto}{John Doe}
%Biography text here.
%\end{IEEEbiographynophoto}

% insert where needed to balance the two columns on the last page with
% biographies
%\newpage

%\begin{IEEEbiographynophoto}{Jane Doe}
%Biography text here.
%\end{IEEEbiographynophoto}

% You can push biographies down or up by placing
% a \vfill before or after them. The appropriate
% use of \vfill depends on what kind of text is
% on the last page and whether or not the columns
% are being equalized.

%\vfill

% Can be used to pull up biographies so that the bottom of the last one
% is flush with the other column.
%\enlargethispage{-5in}

% that's all folks
\end{document}